\documentclass[acmsmall,screen]{acmart}
\AtBeginDocument{%
  }

\usepackage{algorithm}
\usepackage{algorithmic}
\usepackage{graphicx}
\usepackage{textcomp}
\usepackage{xcolor}

\usepackage{makecell}
\usepackage{tabularx}
\usepackage[T1]{fontenc}
\usepackage{comment}
\usepackage{xcolor}
\usepackage{multirow}
\usepackage{array}
\usepackage{soul}
\usepackage{threeparttable}
\usepackage{balance}
\usepackage{diagbox}
\usepackage{subcaption}
\usepackage{microtype} 
\usepackage{calc}

\usepackage{enumitem}
\usepackage{url}
\usepackage{xspace}
\usepackage{float}
\usepackage{stfloats}
\usepackage{adjustbox}
\usepackage{amsmath,amssymb,amsfonts}
\usepackage{algorithmic}
\usepackage{graphicx}
\usepackage{textcomp}
\usepackage{xcolor}
\usepackage{multirow}
\usepackage{mdframed}
\usepackage{tcolorbox}
\usepackage{makecell}
\usepackage{threeparttable}
\usepackage{colortbl}
\usepackage{hhline}
\usepackage{pifont}
\usepackage{enumitem}

\newcolumntype{C}{>{\centering\arraybackslash}X}
\usepackage{pgfplots}
\pgfplotsset{compat=1.16}
\usetikzlibrary{pgfplots.statistics}

\newcommand{\parabf}[1]{\noindent\textbf{#1}}

\newcommand{\zqm}{\textcolor{black}}

\definecolor{ggray}{HTML}{eff0f0}
\definecolor{gggray}{HTML}{E8E8E8}
\definecolor{ggggray}{HTML}{BEBEBE}

\newcommand{\claude}{Claude-Sonnet-4.5\xspace}
\newcommand{\gpt}{GPT-5.1\xspace}
\newcommand{\gemini}{Gemini-3.0-Pro\xspace}
\newcommand{\deepseek}{DeepSeek-V3.1-Terminus\xspace}
\newcommand{\qwencoder}{Qwen3-Coder-480B\xspace}
\newcommand{\qwenfourteen}{Qwen3-14B-Base\xspace}
\newcommand{\qweneight}{Qwen3-8B-Base\xspace}

\newcommand{\qwencodefourteen}{Qwen2.5-Coder-14B-Base\xspace}
\newcommand{\qwencodethreetwo}{Qwen2.5-Coder-32B-Base\xspace}

\newcommand{\ourapproach}{UCD-Training\xspace}
\newcommand{\ourbenchmark}{UnseenCodeBench\xspace}

\definecolor{myyellow}{HTML}{FFF2CC}

\newcounter{finding}
\newcommand{\finding}[1]{\refstepcounter{finding}
 	\vspace{1mm}
	\begin{mdframed}[linecolor=gray!25,roundcorner=12pt,backgroundcolor=myyellow!30,linewidth=3pt,innerleftmargin=2pt, leftmargin=0cm,rightmargin=0cm,topline=false,bottomline=false,rightline = false]
		\textbf{Finding \arabic{finding}:} #1
	\end{mdframed}
	\vspace{1mm}
}

\newcommand{\boxmargin}{1mm}
\tcbuselibrary{most, skins, breakable, theorems}
\newtcolorbox{myboxa}[2][]{
    colback=gray!10!white,
    colframe=black, enhanced,
    attach boxed title to top left={yshift=-2mm,xshift=5mm},
    title=#2,#1
}
\newtcolorbox{myboxb}[2][]{
    boxsep=3pt,
    left = \boxmargin, right = \boxmargin, top = \boxmargin, bottom = \boxmargin,
    title={#2},#1
}
\newtcolorbox{myboxc}{
    colback=gray!15!white,
    arc = 0pt, outer arc = 0pt,
    boxsep=0pt, left = 3pt, right = 0pt, top = 0pt, bottom = 0pt, 
    leftrule=3pt, bottomrule=0pt,toprule=0pt, rightrule=0pt,
    left = \boxmargin, right = \boxmargin, top = \boxmargin, bottom = \boxmargin
}
\newtcolorbox{myboxd}{
    colback=gray!10,%gray background
    colframe=black,% black frame colour
    width=\columnwidth,% Use 8cm total width,
    arc=1mm, auto outer arc,
    boxrule=0.5pt,
}
\usepackage{tikz}

\begin{document}

\title{Unseen-Codebases-Domain Data Synthesis and Training Based on Code Graphs}

\author{Guangsheng Ou\textsuperscript{†}\textsuperscript{\#}}
\affiliation{%
  \institution{Sun Yat-sen University}
  \country{China}}
\email{ougsh3@mail2.sysu.edu.cn}

\author{Qiming Zhang\textsuperscript{†‡}}
\affiliation{%
  \institution{WeChat Pay, Tencent}
  \country{China}}
\email{qmzhangzz@hotmail.com}

\author{Sirong Chen\textsuperscript{\#}}
\affiliation{%
  \institution{WeChat Pay, Tencent}
  \country{China}}
\email{sirongchen49@outlook.com}

\author{Anji Li}
\affiliation{%
  \institution{Sun Yat-sen University}
  \country{China}}
\email{lianj8@mail2.sysu.edu.cn}

\author{Dong Xu}
\affiliation{%
  \institution{Sun Yat-sen University}
  \country{China}}
\email{xudong7@mail2.sysu.edu.cn}

\author{Tiancheng Luo}
\affiliation{%
  \institution{The Chinese University of Hong Kong, Shenzhen}
  \country{China}}
\email{123090401@link.cuhk.edu.cn}

\author{Dekun Dai\textsuperscript{\#}}
\affiliation{%
  \institution{Sun Yat-sen University}
  \country{China}}
\email{daidk@mail2.sysu.edu.cn}

\author{Cuiyun Gao}
\affiliation{%
  \institution{The Chinese University of Hong Kong}
  \country{China}}
\email{cuiyungao@outlook.com}

\author{Long Wang\textsuperscript{*}}
\affiliation{%
  \institution{WeChat Pay, Tencent}
  \country{China}}
\email{oliverlwang@tencent.com}

\author{Jun Zhou}
\affiliation{%
  \institution{WeChat Pay, Tencent}
  \country{China}}
\email{anderszhou@tencent.com}

\author{Mingwei Liu\textsuperscript{*}}
\affiliation{%
  \institution{Sun Yat-sen University}
  \country{China}}
\email{liumw26@mail.sysu.edu.cn}

\author{Zibin Zheng}
\affiliation{%
  \institution{Sun Yat-sen University}
  \country{China}}
\email{zhzibin@mail.sysu.edu.cn}

\thanks{\textsuperscript{\#} Work done during internship at Tencent.}
\thanks{\textsuperscript{†} Equal contribution.}
\thanks{\textsuperscript{‡} Project leader.}
\thanks{\textsuperscript{*} Corresponding author.}

\renewcommand{\shortauthors}{Guangsheng Ou, Qiming Zhang et al.}

\begin{abstract}

In the context of newly release software frameworks, large language models (LLMs) often exhibit poor performance and a high rate of hallucination, as they are not exposed to such environments during training.
Although inference-time augmentation techniques such as retrieval-augmented generation (RAG) can partially mitigate hallucinations, knowledge injection through prompting alone is insufficient to enable models to fully understand the intrinsic relationships among different components of a codebase, or to reason about the correct compositions and apply.
Although explicit knowledge injection can be achieved through post-training, compared with public code domains, unseen codebases typically provide only source code and lack large volumes of high-quality, usage-oriented code that can be directly leveraged as training data. Consequently, existing data synthesis approaches are insufficient to adequately capture unseen codebases usage scenarios when restricted to source code alone.

To address these challenges, we propose \ourapproach, a two-stage training framework for reasoning-aware data synthesis grounded in a code graph constructed from unseen codebases. \ourapproach first parses the source code to build a code graph, then conducts dependency-preserving continued pretraining (CPT) using file-level dependency data, followed by graph-grounded supervised fine-tuning (SFT) on three types of synthesized data augmented with explicit reasoning traces: (1) single-hop relation reasoning data, (2) compositional API reasoning data, and (3) codebase utilization data.
We further introduce a new benchmark, \ourbenchmark, for code generation on unseen codebases and conduct comprehensive experiments across multiple codebases. Results show that \ourapproach consistently outperforms existing baselines, demonstrating strong generality across programming languages, model scales, and architectures—for example, achieving a 25.5\% absolute improvement over RAG with GPT5.1. Compared with existing data synthesis methods, \ourapproach yields performance gains ranging from 7.2\% to 26.1\% of Avg. pass@1 on \ourbenchmark.
Moreover, we simulate realistic enterprise scenarios involving multiple languages and codebases, where a single model needs to adapt to multiple unseen codebases. In these real-world settings, \ourapproach remains effective, achieving a pass@1 of 36.0\% on \ourbenchmark.
\end{abstract}

% \begin{CCSXML}
% <ccs2012>
%    <concept>
%        <concept_id>10011007.10010940.10010971.10010980.10010984</concept_id>
%        <concept_desc>Software and its engineering~Model-driven software engineering</concept_desc>
%        <concept_significance>300</concept_significance>
%        </concept>
%  </ccs2012>
% \end{CCSXML}

% \ccsdesc[300]{Software and its engineering~Model-driven software engineering}

% \keywords{Unseen Codebases Domain, Large Language Models, Continue Pretraining, Supervised Fine-tuning}

\setcopyright{none}

\maketitle

\section{Introduction}
Large language models (LLMs) have demonstrated strong capabilities across a wide range of software engineering tasks, including code generation, completion, translation, and refactoring~\cite{hou2024SEreview,ou2025enhancing,2023SEservey,add1,add2,add3}. These successes, however, largely rely on the availability of abundant training data drawn from mature public code ecosystems (e.g. GitHub)~\cite{chen2021evaluating,jiang2024survey}.

In practice, \textbf{developers frequently work with codebases that are absent from LLM pretraining corpora, such as newly released libraries, domain-specific frameworks, customized enterprise systems, or proprietary third-party components}. We refer to such codebases as \textbf{unseen codebases}. In this setting, LLMs lack the domain-specific knowledge required to correctly understand and \zqm{utilize} the codebase~\cite{zhang2025unseen}, and consequently exhibit severely degraded performance and high hallucination rates~\cite{DomCoder}. 
For instance, prior studies report a performance drop of \textbf{37--57\%} on code generation tasks when models are applied to unseen or private repositories~\cite{zan2022language}. 
\zqm{LLMs} in these settings frequently misuse components, generate invalid compositions, or rely on non-existent functionalities.

A natural first attempt to mitigate this problem is to rely on inference-time augmentation techniques, such as prompt engineering or retrieval-augmented generation (RAG)~\cite{zhang2025llmhallucinations,yang2025empirical,lu2022reacc,shrivastava2023repository,zhou2022docpromptingrag}, which provide access to relevant \zqm{source} code during generation. While such approaches can reduce obvious hallucinations, they do not fundamentally address the underlying issue. Correct code generation in unseen codebases often requires knowledge of how different components are intended to be used together—knowledge that reflects latent usage patterns and conventions learned during training, rather than \zqm{the source code} that can be reliably retrieved alone. As a result, inference-time techniques are \zqm{usually} insufficient to enable robust generalization to unseen codebases.

These observations suggest that \textbf{effectively adapting LLMs to unseen codebases requires explicit knowledge injection through training such as continue pre-training (CPT)~\cite{Zhou2024ContinualLW} and supervised fine-tuning (SFT)~\cite{gunel2020supervised}, rather than relying solely on inference-time augmentation}. However, constructing suitable training data for this purpose is inherently challenging. Unlike public code domains, unseen codebases, \zqm{typically those newly-released codebases}, provide only source code \zqm{implementation}, while high-quality, usage-oriented examples that demonstrate correct component composition are scarce or entirely absent. This data scarcity severely limits the applicability of existing post-training approaches.

Although several data synthesis methods have been proposed to automatically generate training data for LLMs in code generation tasks~\cite{naduaș2025synthetic,wei2023magicoderOSSInstruct,luo2023wizardcoder}, they remain insufficient for handling unseen codebases. 
For example, approaches such as OSS-Instruct~\cite{wei2023magicoderOSSInstruct} typically rely on observable usage patterns from public repositories, implicitly assuming that there are rich real-world examples adequately demonstrating how components should be composed. 
This assumption fails for newly released or private codebases, where code implementation is often the only available signal. 
% Moreover, most existing synthesis pipelines do not consider explicit reasoning traces. Reasoning-centric paradigms have been shown to improve both interpretability and performance, consistently outperforming non-reasoning counterparts~\cite{zhang2025system}. 
% As a result, most conventional data synthesis methods that ignore reasoning processes are ill-suited for post-training modern LLMs intended to handle unseen codebases.

To address these challenges, \textbf{we propose \ourapproach} (\textbf{U}nseen-\textbf{C}odebase-\textbf{D}omain \textbf{Training})\textbf{, a two-stage training framework, consisting of dependency-preserving CPT data and graph-grounded SFT data, designed to inject codebase-specific knowledge into LLMs using only source code implementation from unseen codebases}. Our approach first constructs a \zqm{code graph within the codebases based on code dependency analysis, and constructs dependency-preserving CPT data to capture structural and semantic relationships. We perform CPT to help the model memorize how codebases are implemented.}
% adapt the model to codebase-level characteristics.
We then synthesize usage-oriented SFT training data \zqm{according to three complementary data types}: 
(1) single-hop relation data, 
(2) compositional API data, and 
(3) codebase utilization data. 
\zqm{These data are all augmented with natural reasoning traces to improve LLMs' both interpretability and performance.}

To systematically evaluate this problem, \textbf{we also introduce a new benchmark \ourbenchmark, targeting unseen codebases across two widely used programming languages}, C++ and Python. Extensive experiments demonstrate that \ourapproach consistently outperforms retrieval-based and training-based baselines \zqm{by 7.2\% ot 26.1\% on pass@1}, generalizes across model architectures and scales from 7B to 32B parameters, and remains effective in realistic scenarios \zqm{where one single model needs to adapt to} multiple unseen codebases, achieving a pass@1 of 36.0\% on \ourbenchmark and still surparssing all baselines by 4.7\% to 23.6\%. These results indicate that reasoning-aware data synthesis from \zqm{code implementation alone} can provide a practical and effective solution for adapting LLMs to unseen codebases.

In summary, this paper makes the following contributions:

\begin{itemize}

    \item We propose \textbf{\ourapproach}, a two-stage training framework that injects domain-specific knowledge into LLMs for unseen codebases by performing reasoning-aware data synthesis based on a code graph constructed from source code.
    
    \item We introduce a benchmark, \textbf{\ourbenchmark}, for code generation on unseen codebases across widely used programming languages, providing a standardized evaluation testbed for future research.
    
    \item We conduct \textbf{extensive experiments} to evaluate the effectiveness and generality of \ourapproach, demonstrating that it consistently improves LLMs’ understanding and application of unseen codebases across different languages, model scales, and architectures.
\end{itemize}

% \input{sections/2_background}

% \vspace{-15pt}
\section{Approach}
\label{sec:app}

To address the challenge that existing \zqm{LLMs cannot effectively utilize unseen codebases to solve the requirements,}
we propose \ourapproach, a two-stage \zqm{CPT and SFT training framework. 
} \ourapproach first parses the source code of unseen codebases to build a code graph, \zqm{and then constructs code dependency-preserving CPT data and graph-grounded SFT data. SFT data are synthesized from three views: single-hop relation data, 
compositional API data, and 
codebase utilization data, and augmented with natural reasoning traces.}
These data types enable the model to perform repository-grounded reasoning and generation over the unseen codebase.
Figure \ref{fig:overview} illustrates the overall workflow. The framework operates in the following stages:

\begin{figure*}[b]
    \centering
    \vspace{-10pt}
    \includegraphics[width=1\linewidth]{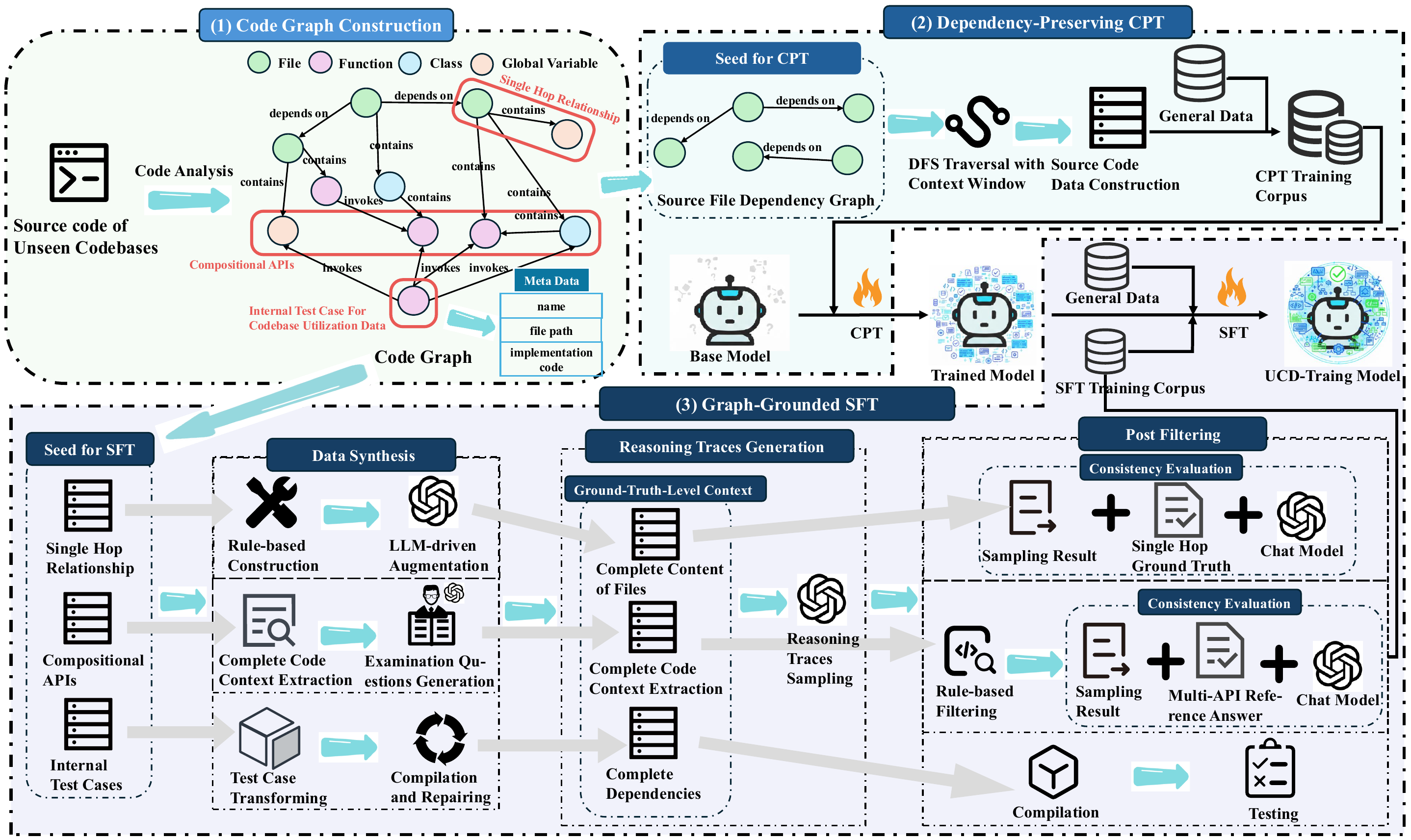}
    \vspace{-20pt}
    \caption{Overview of \ourapproach}
    \label{fig:overview}
    \vspace{-10pt}
\end{figure*}

\begin{enumerate}
    \item \textbf{Code Graph Construction}. We perform program analysis over the source code of the codebase to extract code entities, including files, classes, functions, methods, and global variables as nodes, along with their implementations as their properties. We then parse their dependency, call, include, and other relations as edges, thereby constructing the code graph of the codebase.
    
    \item \textbf{Dependency-Preserving CPT}. The CPT stage begins with data construction. We extract file nodes and their dependency edges from the code graph to obtain a directed acyclic graph that captures file-level dependencies. Starting from nodes with in-degree zero, we perform a depth-first traversal over this graph and \zqm{concatenate the file content according to the resulting file path order}. We then apply a sliding-window truncation strategy \zqm{to the data} to construct training samples that satisfy \zqm{training context length constraints}. This process ensures that code files with direct dependency relationships appear within the same training instance, thereby forming a dependency-preserving source code training corpus. We then mix this corpus with general-domain data and perform CPT.

    \item \textbf{Graph-Grounded SFT}. The SFT stage also begins with data construction. The synthesized SFT data consists of three parts: single-hop relation reasoning data, compositional API reasoning data, and codebase utilization data.
    The single-hop relation reasoning data are constructed from each edge in the code graph and its two endpoint nodes.
    The compositional API reasoning data are synthesized as multiple types of code examination questions based on API combinations extracted from the codebase's internal test cases that cover different functional scenarios of the project. 
    The codebase utilization data is obtained by transforming the codebase's internal test cases into function-generation tasks.
    We then synthesize the reasoning traces for these three types of data by providing ground-truth-level context, thereby completing the data construction for the SFT stage. Finally, the synthetic data are mixed with general-purpose data to perform SFT training.

\end{enumerate}

These three stages form the core of the \ourapproach methodology. Detailed implementation, training dataset construction procedures, and descriptive statistics are provided in the following sections.

\subsection{Code Graph Construction}

In this stage, we parse the source code of unseen codebases to construct a structured code graph. By transforming raw source code into a graph-based representation, we effectively capture the intrinsic relationships among different components of the target codebases, as well as the potential combinations of APIs in application scenarios. This code graph serves as the foundation for training data generation in both the dependency-preserving CPT stage and the graph-grounded SFT stage.

Specifically, we perform program analysis to extract code entities, including files, classes, functions, methods, and global variables as nodes, along with their names, types, implementations, and file locations as properties. 
We then parse their dependency, call, include, and locate relations as edges, thereby constructing the code graph of the codebase as shown in Figure \ref{fig:overview}.

\subsection{Dependency-Preserving CPT}

\subsubsection{Construction of File-level Dependency Data from Code Graph . }
\zqm{We extract only file nodes and inter-file dependency relations from the code graph, resulting in a directed acyclic subgraph, which will be used to construct the pretraining code corpus.} 
Unlike previous works~\cite{guo2024deepseekcoder}, which process project source code by taking a topological ordering for each connected component in the subgraph and then splitting it according to the context window limit to obtain the final file-level dependency data, our approach addresses a key limitation of this strategy. Files that are actually dependent on each other are not necessarily adjacent in topological order. Therefore, when a topological sequence is split due to the context window length limit, some files with dependency relations may not appear in the same training sample, \zqm{leading to an insufficient description of how these files collaborate within the same codebase.}

To address this issue, we perform a depth-first traversal over the directed acyclic subgraph \zqm{as illustrated in Algorithm \ref{alg:DFS Traversal}}. \zqm{We start from nodes with zero in-degree, i.e., files that are not depended on by any other files, and find all depth-first paths in the subgraph}. This strategy ensures that any pair of files with a true dependency relationship appears as adjacent nodes in at least one traversal path. For each path, we generate training samples using a sliding-window strategy. \zqm{Specifically, we fix the left pointer at the beginning of the traversal path and move the right pointer incrementally, until the total length of all file content within the current sequence exceeds the model’s context limit. The concatenated content of the files in the current sequence constructs a CPT training sample. Then, we update the left point location by one step and find the proper right pointer in the same way. We repeat such a process for all traversal paths and construct the dependency-preserving CPT data.}

\begin{algorithm}[t]
\caption{Training Sample Generation via DFS Traversal}
\begin{algorithmic}[1]
\REQUIRE Directed acyclic subgraph $G$, context window limit $L$
\ENSURE Training samples $\mathcal{S}$

\STATE Identify all nodes in $G$ with zero in-degree as start nodes
\FOR{each start node $v$}
    \STATE Perform depth-first traversal to obtain paths $\mathcal{P}$
    \FOR{each path $p \in \mathcal{P}$}
        \STATE $l \leftarrow 1$, $r \leftarrow 1$
        \WHILE{$r \le |p|$}
            \IF{length$(p[l:r]) \le L$}
                \STATE $r \leftarrow r + 1$
            \ELSE
                \STATE Emit $p[l:r-1]$ as a training sample
                \STATE $l \leftarrow r - 1$, $r \leftarrow l$
            \ENDIF
        \ENDWHILE
    \ENDFOR
\ENDFOR
\end{algorithmic}
\label{alg:DFS Traversal}
\end{algorithm}

This method effectively guarantees that files with actual dependency relations will appear at least once adjacently in a single training sequence. As a result, the model can more clearly capture dependency relations among different code elements in the unseen codebase, thereby improving its understanding of the codebase and its ability to reason about cross-file, repository-level dependencies.

\subsubsection{\zqm{CPT Data Corpus Composition and Training.}}
We mix the constructed file-level dependency source code data with general code-relevant data to form the final training corpus for the CPT stage. This is because continual pre-training that uses only domain-specific data can undermine models' original general capability~\cite {wu2025continual}. In such a case, even though the model’s understanding of unseen code codebases improves after CPT, its overall \zqm{coding ability} may actually degrade.

\subsection{Graph-Grounded SFT}

\subsubsection{Construction of Infrastructure and Applications Data with Reasoning Traces from Code Graph}
\zqm{Each data type undergoes three phases: data synthesis, ground-truth–level contextual reasoning traces generation, and post-filtering to eliminate low-quality samples, resulting in high-quality training samples augmented with explicit reasoning traces for understanding and applying unseen codebases.}

\parabf{(1) Single-hop relation reasoning data}
are derived from the code graph to capture fine-grained structural relations among code entities in an unseen codebase, which are essential for understanding its implementation patterns and architectural organization.

\parabf{Data Synthesis. }Specifically, we synthesize single-hop relation data through a two-stage process consisting of rule-based construction followed by LLM-driven augmentation.
For each single-hop relation $A \rightarrow B$, we adopt a rule-based approach to map it into a textual representation of the following form:
\[
[\text{type of } A]\; [\text{name of } A]\; [\text{relation between A and B } ]\; [\text{type of } B]\; [\text{name of } B]
\]
To improve robustness and prevent overfitting to homogeneous positive samples, we further employ an LLM-based approach to diversify each data instance and generate negative samples. Specifically, each original instance is expanded into $N_1$ diversified paraphrased samples and $N_2$ negative samples, where $N_1 = 5$ and $N_2 = 1$.

\parabf{Reasoning Traces Generation. }For positive samples in single-hop relation data, ground-truth–level context refers to the complete contents of the files in which nodes $A$ and $B$ are located. This context explicitly contains the surrounding information of both nodes as well as the associations between them. 
For negative samples, ground-truth–level context consists of a list of all code entity names within the unseen codebases that share the same categories as nodes $A$ and $B$, clearly demonstrating that the fabricated code entity referenced in the query does not exist.
We provide the original instruction together with the constructed contextual information to a \zqm{powerful LLM for} rejection sampling, then record its outputs and reasoning traces.

\parabf{Post Filtering. }Directly assessing the correctness of a model’s reasoning trace is challenging, as the underlying reasoning is often lengthy, information-dense, and difficult to evaluate in isolation. Therefore, we simplify the evaluation by judging the correctness of the model’s output and using it as a proxy for the correctness of the associated reasoning trace.
Specifically, we employ \zqm{a modern} LLM to determine whether the output produced by the reference model is semantically consistent with the constructed ground-truth response and filter out the samples judged as inconsistent. The validated reasoning trace is then embedded into the original data instances, completing the construction of the single-hop relation reasoning data training set.

\parabf{(2) Compositional API Reasoning Data}
are synthesized as multiple types of code examination questions based on coherent API combinations extracted from the codebase’s internal test cases, covering diverse functional scenarios of the project.
Rather than randomly composing APIs, which often results in semantically inconsistent or impractical usage patterns, we leverage internal test cases as high-quality priors. Since such test cases are written by maintainers to validate concrete functionalities, the extracted API combinations naturally reflect realistic usage scenarios and engineering constraints.

\parabf{Data Synthesis. }
To generate high-quality compositional reasoning data, we simulate the process of human-designed examination tasks and formulate a set of explicit task-design principles to guide an LLM in task generation. These principles ensure that the synthesized tasks are structurally well-formed, semantically coherent, and tightly aligned with the target codebase.
Specifically, we constrain the task scope to applications, design principles, and code analysis directly related to the codebase, preventing overly generic or irrelevant tasks. For each task, the model is required to output not only the problem statement, but also a reference answer and explicit grading criteria, forming a complete task–solution–evaluation loop suitable for supervised training. To further increase information density, we additionally require each task to explicitly list its scoring points and align them with corresponding knowledge units, encouraging the model to focus on the codebase’s core concepts and functionalities.

We synthesize three task formats: (1) question–answer tasks targeting conceptual understanding of the codebase design, (2) fill-in-the-blank tasks for context-aware code completion, and (3) programming tasks for end-to-end code generation grounded in the codebase. Task difficulty is explicitly controlled by specifying the number of knowledge points required, providing a concrete and interpretable difficulty signal beyond subjective labels.
For contextual information, we provide not only the implementation code of the selected API combinations, but also precise location information and the complete implementations of all required dependencies. This dependency-closure context enables the model to reason over API interactions holistically, rather than relying solely on abstract signatures.

\parabf{Reasoning Traces Generation. }
Ground-truth–level context for compositional API reasoning data is defined as a subset of the contextual code information used during data synthesis. Since it is difficult to determine which specific pieces of context the model relies on, we perform rejection sampling by providing the problem statement together with the original contextual information and recording the generated responses and reasoning traces. Samples are retained only when the responses are judged to be valid under the provided context.

\parabf{Post Filtering. }
Compositional API reasoning data employ a two-stage filtering strategy.
The first stage is rule-based and applied to the synthesized task descriptions and reference answers. We extract code entity names from task descriptions and parse the reference code using tree-sitter to identify invoked functions and invocation patterns. We then validate (1) whether each function exists in the provided contextual code or is a language built-in, and (2) whether its invocation prefix and argument count match the corresponding function signature. This stage filters out unseen-codebases-domain syntactic errors introduced during task synthesis.

The second stage is applied after reasoning trace generation and combines LLM-based and rule-based validation. A chat model checks semantic consistency between the reference answer and the generated response, while rule-based checks again verify unseen-codebases-domain syntactic correctness. This stage further removes samples with functional or reasoning-level errors, preventing hallucinated behaviors from being propagated to the final training data.

\parabf{(3) Codebase utilization data} 
are obtained by transforming the internal test cases of the unseen codebases. Although such test cases are typically designed to evaluate atomic functionalities, the API invocation patterns they contain and the implicit functional scenarios they encode can substantially enhance a model’s understanding of how project APIs are used and the fundamental application scenarios the project supports.

\begin{figure*}[b]
    \centering
    \vspace{-10pt}
    \includegraphics[width=0.7\linewidth]{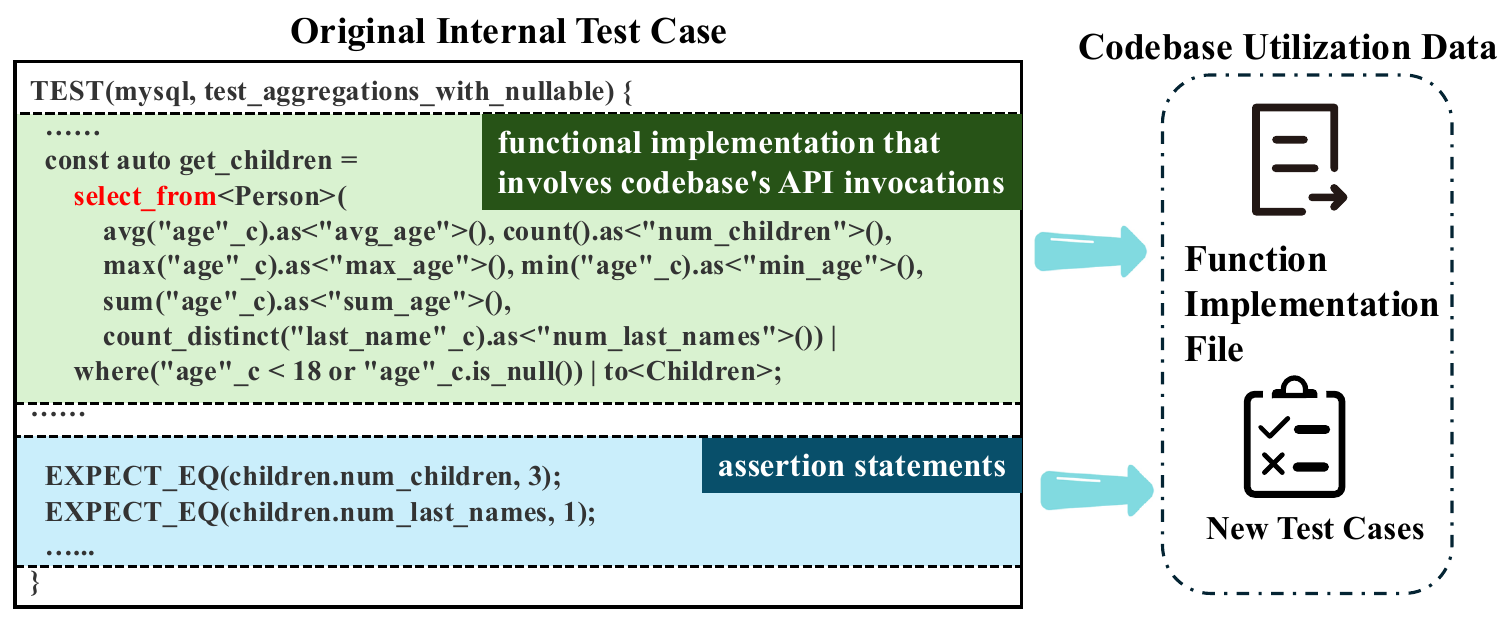}
    \vspace{-10pt}
    \caption{Example for Generating Codebase Utilization Data}
    \label{fig:example_for_generating_api_usage_generation_data}
    \vspace{-10pt}
\end{figure*}

\parabf{Data Synthesis. }Specifically, as shown in Figure \ref{fig:example_for_generating_api_usage_generation_data}, we decompose each internal test case into two components: (1) the functional implementation that involves codebase's API invocations and (2) the assertion statements that verify whether the API behavior matches the expected outcomes. Using an LLM-based few-shot approach, 
% (as illustrated in Figure \ref{fig:prompt_for_generating_api_usage_generation_data})
we perform a test-case decomposition and generate function comments that describe the intended functionality of each implementation. We then apply an iterative compile-and-repair procedure to ensure that both the extracted functional code and the new test cases are syntactically and semantically correct.

\parabf{Reasoning Traces Generation. }Ground-truth-level context of the codebase utilization data is constructed by collecting all API dependencies involved in the functional implementation and corresponding information. 
Specifically, we leverage the code graph to extract functions, data types, and global variables that have explicit call relations with the original test case. For each dependency, we retrieve its concrete implementation, file location, and enclosing namespace.
This dependency-complete context enables the model to reason over which APIs should be invoked, how they interact, and which components are irrelevant to the target functionality. 
We provide the instruction together with this ground-truth dependency information to a \zqm{modern LLM}, record the generated outputs and reasoning traces, and apply rejection sampling

\parabf{Post Filtering. }Similarly, we use the correctness of the generated code as a proxy for the validity of the associated reasoning trace. Each sampled instance is compiled and executed against the test suite, and only those that successfully pass both compilation and testing are retained. The validated reasoning traces are then embedded into the corresponding instances, yielding the final codebase utilization dataset.

\subsubsection{SFT Data Corpus Composition and Training.}
We mix the constructed data with general code relevant data to form the final training corpus for the SFT stage to maintain its general capability as CPT stage.

\section{Benchmark}
Unseen code domains are pervasive in industrial settings. Most enterprises either adapt existing open-source frameworks to meet internal business requirements or develop proprietary third-party libraries to improve development efficiency and security. Although a variety of benchmarks have been introduced to evaluate LLMs across different software engineering dimensions, the majority of these efforts focus primarily on public code scenarios. While prior research has established benchmarks for private codebases \cite{zan2022language,wang2025exploracoder}, it typically concentrates on a single codebase in a single programming language, failing to capture realistic enterprise environments that often involve multiple codebases across multiple programming languages simultaneously.
Therefore, \textbf{we construct an unseen codebases evaluation benchmark, \ourbenchmark}, to assess the effectiveness of \ourapproach.

\parabf{Construction. } 
We choose C++ and Python as our target languages because of their wide adoption and popularity \cite{djurdjev2024popularity}. In particular, C++ is widely used for building system-level infrastructure \cite{stroustrup2013c++}, while Python is extensively used for developing application software \cite{saabith2019python}.
We construct the evaluation benchmark with LLM assistance followed by manual double-checking. Inspired by previous work \cite{zhuo2024bigcodebench,zan2022language} that evaluates a model’s understanding of a specific codebase by testing its ability to use the codebase’s APIs, we first analyze each target codebase to extract all publicly available functions, global variables, and custom data types (e.g., classes). We then manually select combinations of these elements that can jointly implement certain functionalities provided by the codebase, and design corresponding target task scenarios. These are fed into an LLM to generate an evaluation instance consisting of a task description, a reference implementation, and test cases.
Subsequently, we manually check and refine the generated data along several dimensions: the rationality of the task setup, the correctness of the functional description, the executability of the reference implementation and its corresponding test cases, and the consistency between the code and the tests. After this first pass, another author repeats the same quality-checking process to confirm the quality of the evaluation data by double-checking.
The entire construction process is carried out by five engineers, each with 3 to 8 years of programming experience, and in total consumes 100 person-hours.

\begin{table}[tbp]
\caption{Details of Selected Projects for Benchmark Construction.PL: Programming Language, FLD: File-level Dependencies, D.: Dependencies}
\vspace{-8pt}
\centering
\begin{adjustbox}{width=1\linewidth}
        \begin{tabular}{@{}c|c|ccccccc@{}}
            \hline
            \makecell{\textbf{PL}} & \makecell{\textbf{Project Name}} &  \textbf{Created Time} & \textbf{Stars} & \textbf{\#LOC} & \textbf{Number of Files} & \makecell{\textbf{Avg. FLD of File}} & \makecell{\textbf{Avg. D. of Function}} & \makecell{\textbf{Number of Test Cases}}
            \\ 
            \hline
            \multirow{3}{*}{C++} 
            & sqlgen\cite{sqlgen} & 2025.3  & 131 & 14,164  & 213 & 3.5 & 5.9 & 60 \\ \cline{2-9}
            & reaction\cite{reaction} & 2025.3  & 618 & 5,720  & 26 & 2.7 & 4.0 & 64  \\ \cline{2-9}
            & Hexi\cite{hexi}  & 2025.3 & 279 & 6,108 & 29 & 3.1 & 2.0 & 64 \\ 
            \hline
            python & LEANN\cite{leann} & 2025.6  & 8,038 & 7,491  & 14 & 1.6 & 5.1 & 64 \\ 
            \hline

        \end{tabular}
        \end{adjustbox}
\label{tab:details_of_selected_projects}
\vspace{-5pt}
\end{table}

\parabf{Quality Control Mechanism. }
To ensure that the selected target codebases have not appeared in the training corpora of the evaluated models, thus achieve the effect of truly unseen codebases evaluation, we adopt a two-stage verification process.
In the first stage, we ensure that the creation time of each selected project is later than the knowledge cutoff of the models. Based on the knowledge cutoff dates of all the models we evaluate, we take their union and obtain a time threshold \(T\), which we currently set to \texttt{2025-03-01}. By querying GitHub with \texttt{created:>=2025-03-01}, we crawl all projects that satisfy this constraint, and then apply an initial filtering mechanism with \texttt{stars:>100} and \texttt{size:>1000} to eliminate projects that are too small in scale, have low recognition, or lack sufficient activity as previous works\cite{han2019characterization,ou2025repository}, thereby ensuring the quality of the selected projects. Finally, we manually review and select the final target codebases, covering a diverse range of domains, including database systems, binary data stream processing, programming frameworks, and high-performance retrieval engines. Detailed information of selected codebases is shown in Table \ref{tab:details_of_selected_projects}.
In the second stage, we evaluate the performance of the base models on our constructed evaluation set. After constructing the evaluation benchmark, we measure the performance of the base models without any external knowledge augmentation. This allows us to assess how well existing models master unseen-codebases-domain knowledge before any unseen-codebases-domain post-training. As shown in Table \ref{tab:rq1_result}, all selected SOTA models with only base model exhibit a dramatic performance drop on our constructed evaluation set compared to existing public benchmarks, which further indicates that the chosen codebases did not appear in the training corpora of these models. This demonstrates that our constructed evaluation benchmark can effectively measure models’ real performance in unseen-codebases-domain scenarios.
\section{Experiment Setup}

We evaluate the effectiveness, usefulness, and generalization of \ourapproach by answering the following seven research questions (RQs):

\begin{itemize}[left=0em]
    \item \textbf{RQ1 (Performance of \ourapproach)}: How does \ourapproach perform compared to other methods?

    \item \textbf{RQ2 (Ablation Study)}: How does each component of \ourapproach contribute to the final performance (including different categories of training data and filtering mechanisms)?

    \item \textbf{RQ3 (Generality of \ourapproach)}: How generalizable is \ourapproach across different programming languages, model sizes, and model families?
    
    \item \textbf{RQ4 (Application in Real-World Enterprise Scenarios)}: How does \ourapproach perform in simulated real-world enterprise scenarios, i.e., can a single model simultaneously handle multiple unseen codebases across multiple languages?
    
    \item \textbf{RQ5 (Configuration)}: How do different configurations of hyperparameters in \ourapproach affect the results?

\end{itemize}

\subsection{Baseline}
We evaluate the performance of \ourapproach with two types of methods:
(1) Explicit knowledge injection through training: OSS-Instruct~\cite{wei2023magicoderOSSInstruct} and Cotton~\cite{yang2024chainCoT}, which fine-tune the model with synthetic data to enhance its coding capabilities;
(2) Inference-time augmentation: RAG~\cite{zhang2025llmhallucinations}, which uses retrieval to mitigate the model’s hallucination phenomenon.

\parabf{OSS-Instruct}~\cite{wei2023magicoderOSSInstruct} is a proven and influential program synthesis approach, distinguished by its effectiveness in leveraging synthetic data to fine-tune small-parameter models. It serves as a mature, robust baseline in code generation and is widely adopted in the field of code data synthesis. We compare this method with our proposed data synthesis approach to demonstrate the effectiveness of our data synthesis pipeline.

\parabf{COTTON}~\cite{yang2024chainCoT} is a Chain-of-Thought (CoT) based program synthesis approach specifically designed for lightweight language models. It automatically generates high-quality CoTs without relying on manually crafted rationales to fine-tune models so as to improve the LLMs' reasoning capability, thereby enhancing their code generation ability. We compare the CoT synthesis method in COTTON with our proposed reasoning-process synthesis method to validate the effectiveness of our approach for generating reasoning traces.

\parabf{RAG}~\cite{zhang2025llmhallucinations} retrieves code snippets from the project that are similar to the current task and incorporates them into the input prompt to help LLMs better understand the requirements and gain awareness of specific factual knowledge and project contexts.

\subsection{Studied LLMs}

As shown in Table \ref{tab:studied_llms}, For the RAG-based baselines, we select several state-of-the-art models, covering both general-purpose LLMs and code-oriented LLMs, as well as open-source and closed-source models. We also ensure that the knowledge cutoff of each chosen model is earlier than the creation time of the projects used to build our evaluation benchmark. Although \deepseek and \qwencoder do not publicly disclose their exact knowledge cutoff dates, based on LLMs' training timelines, the association between release time and knowledge cutoff of other models, and their base-model performance on our benchmark, we reasonably infer that the selected projects are not included in the training corpora of these two models either.

For \ourapproach and the training-based baselines, following previous work~\cite{ma2025unitcoder,tao2026swe}, we adopt four models from the Qwen family, which are widely used in post-training scenarios: \qweneight, \qwenfourteen, \qwencodefourteen, and \qwencodethreetwo. This selection covers multiple parameter scales and different model types, enabling a comprehensive evaluation of our method under diverse model capacities and architectures.

\begin{table}[tbp]
\caption{Studied LLMs}
\vspace{-8pt}
\centering
\begin{adjustbox}{width=0.95\linewidth}
\begin{tabular}{@{}c|c|ccccc@{}}
\hline
\textbf{Model Type} & \textbf{Model Name} & \textbf{Open-source} & \textbf{Reasoning} & \textbf{Time} & \textbf{Knowledge Cutoff} & \textbf{Size} \\ \hline

\rowcolor{gray!10} \multicolumn{7}{c}{\textit{Models for RAG}} \\ \hline
\multirow{4}{*}{General LLM} 
& \claude\cite{claude4.5}       & \ding{55}  & \checkmark & 2025.9  & 2025.1 & -  \\ 
& \gpt\cite{gpt-5-1}          & \ding{55}  & \checkmark & 2025.11 & 2024.9 & -  \\ 
& \gemini\cite{gemini-3-pro}       & \ding{55}  & \checkmark & 2025.11  & 2025.1 & -  \\ 
& \deepseek\cite{deepseek-v3.1-teminus}     & \checkmark & \checkmark & 2025.9  & - & 671B  \\ \hline
\multirow{1}{*}{Code LLM} 
& \qwencoder\cite{qwen3technicalreport}     & \checkmark & \ding{55} & 2025.7 & - & 480B  \\  \hline

\rowcolor{gray!10} \multicolumn{7}{c}{\textit{Models for Training}} \\ \hline
\multirow{2}{*}{General LLM} 
& \qweneight\cite{qwen3technicalreport}       & \checkmark  & \ding{55} & 2025.4  & - & 8B  \\ 
& \qwenfourteen\cite{qwen3technicalreport}     & \checkmark & \ding{55} & 2025.4  & - & 14B  \\ \hline
\multirow{2}{*}{Code LLM} 
& \qwencodefourteen\cite{hui2024qwen2coder}     & \checkmark & \ding{55} & 2024.11 & - & 14B  \\  
& \qwencodethreetwo\cite{hui2024qwen2coder}       & \checkmark  & \ding{55} & 2024.11  & - & 32B  \\ \hline

\end{tabular}
\end{adjustbox}
\label{tab:studied_llms}
\vspace{-10pt}
\end{table}

\subsection{Metrics}
Following the previous methods~\cite{wang2022compilable,chen2021evaluating}, we employ the execution-based metrics \textit{Compilation@k} and \textit{Pass@k} to assess the quality of generated code through the execution status of the code. 

\begin{itemize}[left=0em]

\item \textbf{\textit{Compilation@k}}: Compilation@k 
% as expressed in Equation~\ref{eq:CA}
is a crucial evaluation metric that measures the proportion of translated code snippets that compiles successfully without errors within $k$ rounds, directly reflecting the syntactic and structural correctness of the translated code. $C(i, k) = 1$ if the $i^{th}$ code sample compiles successfully within $k$ attempts; otherwise $C(i, k) = 0$.

\item \textbf{\textit{Pass@k}}: This widely-used metric~\cite{chen2021evaluating} calculates the percentage of tasks correctly solved based on $k$ generated code samples per task. 
A task is considered solved if at least one of the generated code samples passes all the corresponding test cases. 

\end{itemize}

\subsection{Implementation Details}

\parabf{Code Parsing.}
During the code graph construction stage, C++ codebases are parsed and compiled using Clang~\cite{clang}, while Python codebases are statically analyzed using Jedi~\cite{jedi} and the built-in AST module. For rule-based post-filtering of compositional API usage reasoning data, we adopt Tree-sitter~\cite{tree-sitter} to perform fine-grained syntactic analysis.

\parabf{Data Construction.}
For the dependency-preserving CPT stage, we incorporate general-domain data from Mixture-of-Thought~\cite{openr1}, retaining only categories relevant to code-related capabilities, including math,code, and scientist reasoning.
For the graph-grounded SFT stage, we use AM-distilled-data\cite{zhao202514millionopensourcedistilled} as general-domain supervision, which includes not only code-related data but also instruction-following data.
The modern model used during graph-grounded SFT is DeepSeek-v3.1-Terminus\cite{deepseek-v3.1-teminus} in reasoning mode, while the model used for consistency evaluation is DeepSeek-v3.1-Terminus in chat mode.
Due to variations in programming languages and repository scales, 
the sizes of the synthesized datasets differ across codebases: 35k samples for sqlgen, 20k samples for Reaction and Hexi, and 6k samples for Leann. For a fair comparison, training-based baselines are trained using the same data scales.

\parabf{Training Setup.}
Due to variations in programming languages and codebases scales, the number of epochs used in the dependency-preserving CPT stage differs across codebases. For C++ codebases, the CPT stage is conducted for 1 epoch on sqlgen and for 2 epochs on reaction and hexi. For the Python codebase leann, CPT is performed for 1 epoch.
In the graph-grounded SFT stage, all codebases are trained for 3 epochs.
For other hyperparameters, both CPT and SFT stages employ full-parameter fine-tuning with a learning rate of $5.0e^{-5}$, a warmup ratio of 0.1, a context window length of 32,768 tokens, and a gradient accumulation step of 1.
All experiments are conducted on 32 NVIDIA H20 GPUs. The complete configuration and training details are available on GitHub.

\section{Evaluation}
\label{sec:result}

\subsection{RQ1: Performance of \ourapproach}
\label{sec:rq1}

\parabf{Design.}
To systematically evaluate the effectiveness of \ourapproach in unseen codebase scenarios, we compare it against two representative categories of methods: inference-time augmentation and training-time augmentation.
For inference-time augmentation, we select Retrieval-Augmented Generation (RAG) as a representative baseline and five state-of-the-art models as shown in Table \ref{tab:studied_llms} as base model.
For training-time augmentation, we consider two widely adopted data synthesis and fine-tuning approaches: OSS-Instruct and COTTON and select the same base model as \ourapproach: \qweneight and \qwenfourteen 
We adopt compilation@1 and pass@1 for C++ codebases and pass@1 for Python codebase LEANN, to assess code generation performance from both syntactic correctness and semantic correctness perspectives in unseen codebases domain.

\begin{table}[tbp]
\caption{Comparison of Ours and Baselines}
\vspace{-8pt}
\centering
\begin{adjustbox}{width=1\linewidth}
        \begin{tabular}{@{}c|c|cc|cc|cc|c|c@{}}
            \hline
            
        \multirow{2}{*}{\textbf{Tech.}}  & \multirow{2}{*}{\textbf{Base Model}} 
          &  \multicolumn{2}{c|}{\textbf{sqlgen}} &  \multicolumn{2}{c|}{\textbf{reaction}} &  \multicolumn{2}{c|}{\textbf{Hexi}} & \multicolumn{1}{c|}{\textbf{LEANN}} & \textbf{\ourbenchmark}\\ \cline{3-10}
  
          & & compilation@1 & \multicolumn{1}{c|}{pass@1} & compilation@1 & \multicolumn{1}{c|}{pass@1} & compilation@1 & \multicolumn{1}{c|}{pass@1} & pass@1 & Avg. pass@1 \\ \hline
          
            % \textbf{Tech.} & \textbf{Model} & \textbf{Compilation@1} & \textbf{Pass@1} & \textbf{DSR@1} & \textbf{RR} & \textbf{CodeBLEU} & \textbf{\astmatchscore} \\ \hline
            \multirow{5}{*}{Base Model} 
            & \claude     & 1.8\% & 1.0\% & 3.8\% & 0.5\%  & 16.1\% & 14.1\% & 20.8\%  & 9.1\% \\ 
            & \gpt        & 0.0\% & 0.0\%  & 2.8\% & 1.9\% & 11.9\% & 11.1\% & 10.6\%  & 5.9\%\\ 
            & \gemini     & 3.3\% & 1.7\% & 3.0\% & 0\% & 20.2\% & 17.0\% & 4.7\%  & 5.9\% \\ 
            & \deepseek   & 0.0\% & 0.0\% & 3.3\% & 2.0\% & 16.3\% & 14.5\% & 10.3\% & 6.7\%  \\ 
            & \qwencoder  & 1.3\% & 0.7\% & 3.1\% & 1.6\% & 20.5\%  & 16.7\% & 12.7\% & 7.9\% \\ 
            \hline
            \multirow{5}{*}{RAG~\cite{zhang2025llmhallucinations}} 
            & \claude     & 11.2\% & 9.5\%  & 11.6\% & 2.2\% & 44.2\% & 39.4\% & 34.5\% & 21.4\%  \\ 
            & \gpt        & 4.2\% & 4.2\% & 14.7\% & 5.5\% & 31.6\% & 28.8\% & 13.4\% & 13.0\%  \\ 
            & \gemini     & 8.3\% & 6.5\% & 11.3\% & 4.5\% & 43.6\% & 38.9\% & 9.1\%  & 14.8\% \\ 
            & \deepseek   & 4.5\%  & 4.2\% & 13.4\% & 4.2\% & 29.8\% & 25.8\% & 18.8\% & 13.3\%  \\ 
            & \qwencoder  & 7.8\% & 7.7\% & 5.9\% & 1.7\% & 23.6\% & 20.0\% &  14.4\% & 11.0\% \\ 
            \hline
            \multirow{2}{*}{OSSInstruct~\cite{wei2023magicoderOSSInstruct}}  
            & \qweneight & 1.2\% & 0.0\%    & 5.8\% & 2.5\%    & 21.3\% & 18.3\%   & 17.0\% & 9.5\%   \\ 
            & \qwenfourteen & 0.7\% & 0.2\% & 2.2\% & 1.7\%    & 24.5\% & 20.3\%   & 27.5\%  & 12.4\% \\ 
            
            \hline
            \multirow{2}{*}{COTTON~\cite{yang2024chainCoT}}  
            & \qweneight    & 18.0\% & 11.2\%   & 30.2\%  & 17.0\%  & 43.9\%  & 35.0\%   & 42.7\% & 26.5\%  \\ 
            & \qwenfourteen & 20.5\% & 16.2\% & 33.3\%  & 18.8\%   & 44.8\%  & 39.1\%  &  50.9\% & 31.3\%  \\ 
            
            \hline
            \multirow{3}{*}{\textbf{\ourapproach}}  
            & {\qweneight} & \textbf{22.8\%} & \textbf{19.1\%}  & \textbf{34.5\%}& \textbf{20.0\%} & \textbf{45.8\%} & \textbf{38.8\%} &  \textbf{47.8\%} &  \textbf{31.5\%}  \\ 
            & {\qwenfourteen} & \textbf{29.3\%} & \textbf{22.2\%} & \textbf{42.0\%} & \textbf{29.1\%} & \textbf{54.2\%} & \textbf{47.2\%} &  \textbf{55.2\%}  &  \textbf{38.5\%} \\ 
            & {\qwencodefourteen} & \textbf{23.3\%} & \textbf{18.2\%} & \textbf{48.8\%} & \textbf{27.3\%} & \textbf{52.5\%} & \textbf{45.3\%} & \textbf{48.1\%}   &  \textbf{34.7\%}\\
            
            \hline
            {\makecell{\textbf{\ourapproach}\\\textbf{(combined)}}}  
            & {\qwenfourteen} & \textbf{23.3\%} & \textbf{19.8\%} & \textbf{44.4\%} & \textbf{27.2\%} & \textbf{54.7\%} & \textbf{46.1\%} &  \textbf{50.9\%} &  \textbf{36.0\%} \\ 

            \hline

        \end{tabular}
        \end{adjustbox}
\label{tab:rq1_result}
\vspace{-10pt}
\end{table}

\begin{figure*}[b]
    \centering
    \vspace{-10pt}
    \includegraphics[width=0.95\linewidth]{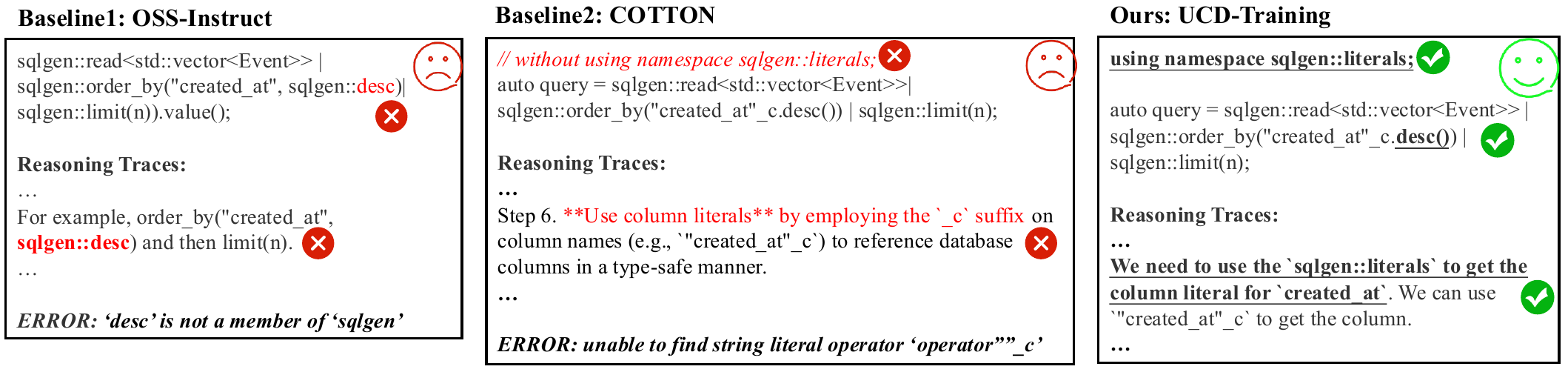}
    \vspace{-10pt}
    \caption{A Comparative Example Illustrating the Outputs on the Same Problem Using Baselines and \ourapproach. The example is the task with ID {limit\_recent\_events} from the sqlgen in \ourbenchmark. The complete model outputs can be found on GitHub.}
    \label{fig:comparison_example}
    \vspace{-10pt}
\end{figure*}

\parabf{Result.}
Table \ref{tab:rq1_result} presents the overall performance comparison between \ourapproach and baseline methods across different unseen codebases. The results reveal several notable findings.

\parabf{First, compared with inference-time augmentation methods,} models trained with \ourapproach significantly outperform RAG-based state-of-the-art baselines on \ourbenchmark, even at the 8B parameter scale. Specifically, our 8B model achieves a pass@1 score of 31.5\%, exceeding the best-performing RAG baseline (\claude) by 10.1\%.
While RAG partially alleviates the difficulty faced by LLMs without domain-specific adaptation in understanding and using unseen codebases, the resulting improvements remain limited.
In particular, as codebases grow in scale and structural complexity, the effectiveness of RAG diminishes substantially. On the relatively simple Hexi codebase, the performance of an 8B model trained with \ourapproach is comparable to that of the best RAG-based model. However, on the most complex and largest codebase, sqlgen, in \ourbenchmark, the gap becomes pronounced. The best RAG-based model is \claude, achieving only 17.2\% compilation@1 and 9.5\% pass@1, while under \ourapproach, even an 8B model reaches 22.8\% compilation@1 and 19.1\% pass@1. These results indicate that merely retrieving and injecting local code snippets is insufficient for enabling models to capture implicit inter-component dependencies and usage conventions, highlighting the necessity of post-training–based knowledge injection for unseen codebases.

\parabf{Second, compared with other explicit knowledge injection through training methods,} \ourapproach consistently yields superior performance under the same model scale across all codebases. It is worth noting that on larger and more complex codebases, the performance gains brought by \ourapproach become even more pronounced. In such settings, models trained with \ourapproach at 8B parameters even outperform 14B-parameter models trained using baseline methods. This advantage stems from two key factors:
(1) \ourapproach synthesizes richer architectural and usage-oriented data grounded in unseen codebases, substantially enhancing the model’s understanding and application of unseen codebases; and
(2) the reasoning traces generated by our method are more realistic and natural than generic Chain-of-Thought data, providing models with stronger reasoning capabilities and enabling them to solve problems effectively in more complex scenarios.

We further compare \ourapproach and existing training-based data synthesis methods through a concrete case study.
As shown in Figure~\ref{fig:comparison_example}, \ourapproach achieves a pass@1 score of 90\% on this task, while all other baselines fail completely with 0\%. OSS-Instruct struggles in this setting because it relies on pre-existing usage-oriented code and cannot adequately cover realistic usage scenarios when only source code is available, leading to hallucinated or invalid API invocations. Although COTTON correctly identifies all required dependencies, the synthesized CoT traces often diverge from the model’s intrinsic reasoning patterns, resulting that models trained on such traces suffer from degraded reasoning robustness and exhibit poor performance in complex scenarios. In contrast, \ourapproach not only accurately identifies all necessary dependencies but also correctly infers and applies the appropriate usage patterns for each dependency, enabling reliable end-to-end code generation.

\finding{\ourapproach effectively injects repository-level structural knowledge and reasoning processes during training, achieving a peak performance of 38.5\% on \ourbenchmark. This result surpasses the best-performing RAG baseline by 10.1\% and the strongest training-based baseline by 7.2\%.}

% \vspace{-10pt}
\subsection{RQ2: Ablation Study}
\label{sec:rq2}

\parabf{Design.}
To gain a deeper understanding of how each key design component of \ourapproach contributes to the overall performance, we conduct a systematic ablation study along two dimensions: training data sources, and data quality control mechanisms. All ablation experiments are performed using the Qwen3-14B-base model and evaluated across all unseen codebases and metrics with comprehensive experiments, ensuring fair and comparable results.

Specifically, we design our ablations from the following perspectives:
\textbf{(1) Ablation on SFT Data Types.}
To verify the complementarity of the three types of graph-grounded SFT data, we selectively remove each data type:
\textit{w/o codebase utilization data}, removing codebase utilization data to assess the importance of realistic usage patterns and end-to-end generation supervision;
\textit{w/o single-hop realtion reasoning data}, removing single-hop relation reasoning data to evaluate the role of fine-grained structural modeling and local dependency learning;
\textit{w/o compositional API reasoning data}, removing compositional API reasoning data to measure the contribution of synthesized unseen-codebase API usage data in the SFT stage.
\textbf{(2) Ablation on General-Domain Data in SFT.}
We further introduce a \textit{with only general data} setting, where all unseen-codebase–specific data are removed. This setup aims to demonstrate that general-domain data mainly serves to preserve the model’s generic code capabilities, while providing little benefit to domain-specific understanding and usage in unseen codebases.
\textbf{(3) Ablation on Data Filtering Mechanisms.}
To analyze the impact of our two-stage filtering strategy on synthesized data quality and downstream performance, we independently remove:
\textit{w/o filter of problem and reference answers}, which eliminates filtering at the question–answer pair level and \textit{w/o filter of reasoning content}, which removes filtering applied to reasoning traces and final responses.

\parabf{Result.}
Table \ref{tab:ablation_result} reports the performance of the full \ourapproach and its ablated variants across different tasks. The results clearly demonstrate the critical contributions of each component.

\begin{table}[tbp]
\caption{Ablation Study on base model \qwenfourteen. CUD: codebase utilization data, SHRRD: single-hop relation reasoning data, CARD: compositional API reasoning data}
\vspace{-8pt}
\centering
\begin{adjustbox}{width=1\linewidth}
\begin{tabular}{@{}c|cc|cc|cc|c@{}}
\hline
\multirow{2}{*}{\textbf{Tech.}} 
& \multicolumn{2}{c|}{\textbf{sqlgen}} 
& \multicolumn{2}{c|}{\textbf{reaction}} 
& \multicolumn{2}{c|}{\textbf{Hexi}} 
& \multicolumn{1}{c}{\textbf{LEANN}} \\ \cline{2-8}

& compilation@1 & pass@1 
& compilation@1 & pass@1 
& compilation@1 & pass@1 
& pass@1 \\ \hline

\ourapproach
& 29.3\% & 22.2\%
& 42.0\% & 29.1\%
& 54.2\% & 47.2\%
& 55.2\% \\ \cline{1-1}

w/o CUD
& 23.8\%~(↓5.5) & 21.7\%~(↓0.5)
& 33.1\%~(↓8.9) & 21.6\%~(↓7.5)
& 53.4\%~(↓0.8) & 46.1\%~(↓1.1)
& 49.8\%~(↓5.4) \\ \cline{1-1}

\makecell{w/o SHRRD}
& 27.5\%~(↓1.8) & 21.3\%~(↓0.9)
& 33.0\%~(↓9.0) & 19.8\%~(↓9.3)
& 50.6\%~(↓3.6) & 44.2\%~(↓3.0)
& 54.6\%~(↓0.6) \\ \cline{1-1}

\makecell{w/o CARD}
& 12.7\%~(↓16.6) & 9.8\%~(↓12.4)
& 37.8\%~(↓4.2) & 24.2\%~(↓4.9)
& 34.4\%~(↓19.8) & 30.8\%~(↓16.4)
& 25.5\%~(↓29.7) \\ \cline{1-1}

with only general data
& 0.5\%~(↓28.8) & 0.5\%~(↓21.7)
& 0.5\%~(↓41.5) & 0.3\%~(↓28.8)
& 28.8\%~(↓25.4) & 24.4\%~(↓22.8)
& 10.5\%~(↓44.7) \\ \cline{1-1}

\makecell{w/o filter of problem and reference\\ answers of CARD}
& 18.2\%~(↓11.1) & 15.2\%~(↓7.0)
& 40.0\%~(↓2.0) & 27.2\%~(↓1.9)
& 52.5\%~(↓1.7) & 44.7\%~(↓2.5)
& 39.1\%~(↓16.1) \\ \cline{1-1}

\makecell{w/o filter of reasoning\\ content of CARD}
& 25.3\%~(↓4.0) & 19.0\%~(↓3.2)
& 42.0\%~(+0.0) & 28.0\%~(↓1.1)
& 55.9\%~({↑1.7}) & 46.7\%~(↓0.5)
& 50.2\%~(↓5.0) \\ \cline{1-1}

\hline
\end{tabular}
\end{adjustbox}
\label{tab:ablation_result}
\vspace{-15pt}
\end{table}

\textbf{First, for ablation on SFT data types}, 
the result shows that all three types of SFT data are indispensable for improving performance on unseen codebases. Removing any single data type consistently leads to noticeable degradation, indicating that they provide complementary benefits for understanding and utilizing unseen codebases. Among them, compositional API reasoning data is the most critical: its removal causes a dramatic performance drop—for example, on LEANN, pass@1 decreases from 55.2\% to 25.5\%, a decline of 29.7\%. This result underscores a key strength of our approach: by automatically synthesizing multi-API usage tasks, we effectively mitigate the scarcity of high-quality, usage-oriented code beyond implementation-level source code, substantially enhancing the model’s ability to solve tasks on the target codebase.

\textbf{Second, for ablation on general-domain data in SFT}, general-domain data mainly serves to preserve the model’s generic capabilities and contributes little to domain-specific generalization on its own. Under the with only general data setting, the model achieves near-zero pass@1 on sqlgen and reaction with 0.5\% and 0.3\% respectively, and only 10.5\% on LEANN. These results closely mirror the behavior of base models in RQ1, further confirming that without explicit private-domain knowledge injection, LLMs struggle to generalize to unseen codebases.
Notably, the model performs better on Hexi than all base models reported in RQ1, even under this setting. This observation indicates that dependency-preserving CPT alone already injects meaningful private-domain knowledge, enabling the model to acquire a non-trivial understanding of the unseen codebase structure, despite the absence of graph-grounded SFT data.

\textbf{Finally, for ablation on data filtering mechanisms}, the strict data filtering mechanism plays a crucial role in ensuring stable performance improvements. Removing any stage of the filtering pipeline leads to consistent performance degradation, indicating that each filtering stage contributes non-trivially to data quality. Moreover, compared with removing the filtering of reasoning content, removing the filtering of problem statements and reference answers results in a substantially larger performance drop.
It is also worth noting that removing problem–answer filtering leads to substantial performance degradation, with pass@1 drops of 7.0\% and 16.1\% on sqlgen and LEANN, respectively—significantly larger than those observed on reaction (1.9\%) and Hexi (2.5\%). This discrepancy stems from the larger scale and higher structural complexity of these unseen codebases. Without post-training on the target codebase, the modern model is more likely to generate problems and reference answers that misuse or misinterpret repository-specific APIs and design assumptions. When filtering is removed, such hallucinations are propagated into the synthesized data, ultimately impairing the model’s understanding and utilization of the target codebase. Consequently, post-filtering becomes increasingly critical as codebase complexity grows.
% \vspace{-10pt}
\finding{Each component of \ourapproach contributes positively to the performance of model. Among them, the compositional API reasoning data in the SFT stage has the largest impact, with a performance drop of up to 29.7\% in w/o CARD setting.}
% \vspace{-20pt}

% \vspace{-10pt}
\subsection{RQ3: Generality of \ourapproach}
\label{sec:rq3}

\parabf{Design.}
To demonstrate the robustness and generality of \ourapproach across diverse settings, we evaluate its performance under variations along three key dimensions: programming languages, base model types, and model scales.
First, to assess language generalization, we conduct experiments on unseen codebases written in two widely used programming languages: Python and C++.
Second, to examine model-type robustness, we evaluate both general LLMs and code LLMs as base models.
Third, to analyze scalability with respect to model size, we consider models with 8B, 14B, and 32B parameters.
For the 32B setting, we adopt \qwencodethreetwo as the base model, since Qwen3 does not currently provide a released base model at this scale.
Due to computational resource constraints, experiments with 32B-scale models are performed on one representative codebase per programming language.

\parabf{Result.}
The experimental results demonstrate that \ourapproach can be applied effectively in diverse settings.
As shown in Table 3, \ourapproach achieves consistent and significant improvements on both C++ and Python codebases, indicating its robustness across programming languages. Moreover, Table 4 shows that, in the Python setting, each key design component of \ourapproach continues to contribute positively to enhancing the model’s understanding and usage of unseen codebases, further validating the necessity of our overall design.
In addition, \ourapproach is effective for both general LLMs and code LLMs. As illustrated in Table 3, \qwencodefourteen consistently outperforms all baseline methods across all evaluated codebases, demonstrating that our method is model-agnostic and does not rely on a specific class of base models.
Importantly, the effectiveness of \ourapproach scales favorably with model size. As shown in Figure 4, when increasing the model size from 8B to 14B parameters, pass@1 improves from 20.0\% to 29.1\% on one codebase and from 47.8\% to 55.2\% on another. When further scaling from 14B to 32B parameters, performance continues to increase, from 27.3\% to 33.8\% and from 48.1\% to 58.8\%, respectively. These results indicate the practical potential of \ourapproach in real-world scenarios with larger base models when given sufficient computational resources.

\finding{\ourapproach consistently enhances models’ understanding and application of target codebases across different programming languages, model types, and model sizes. Moreover, its strong scaling behavior highlights its practical potential for deployment in real-world, industrial-scale software development scenarios.}

\begin{figure*}[b]
    \centering
    \vspace{-10pt}
    \begin{subfigure}{0.4\textwidth}
        \centering
        \includegraphics[width=\linewidth]{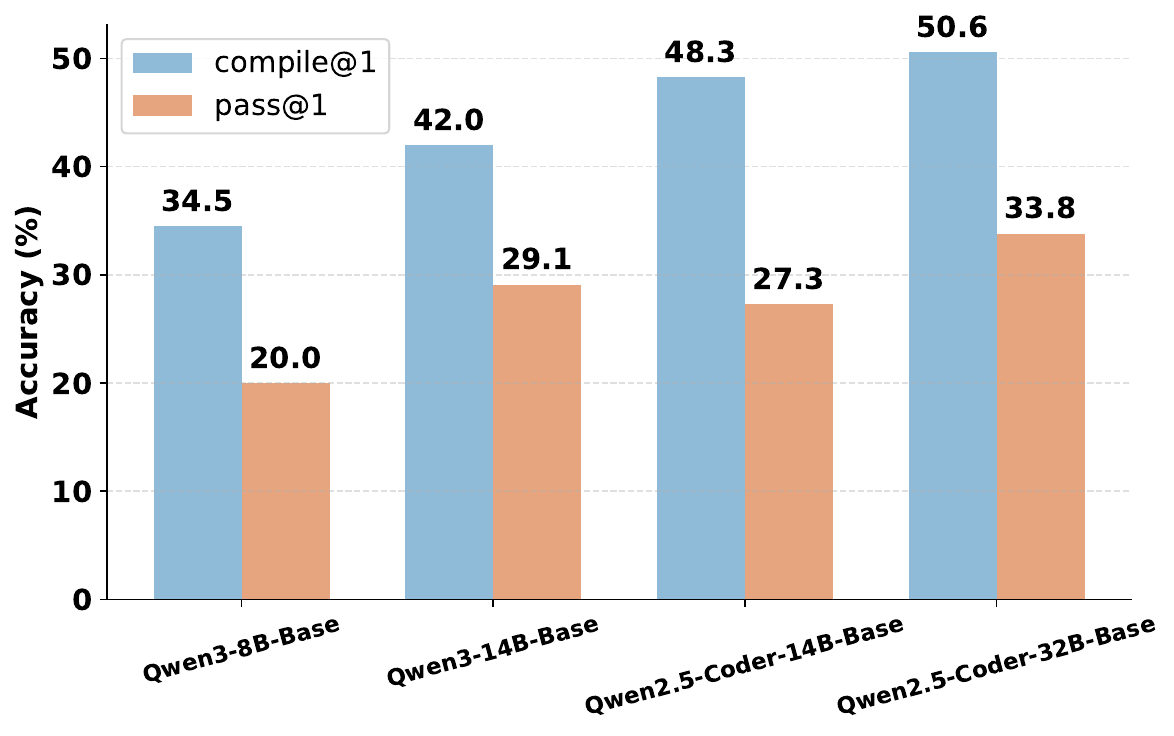}
        \caption{reaction}
        \label{fig:reaction}
    \end{subfigure}
    \hspace{0.05\textwidth}
    \begin{subfigure}{0.4\textwidth}
        \centering
        \includegraphics[width=\linewidth]{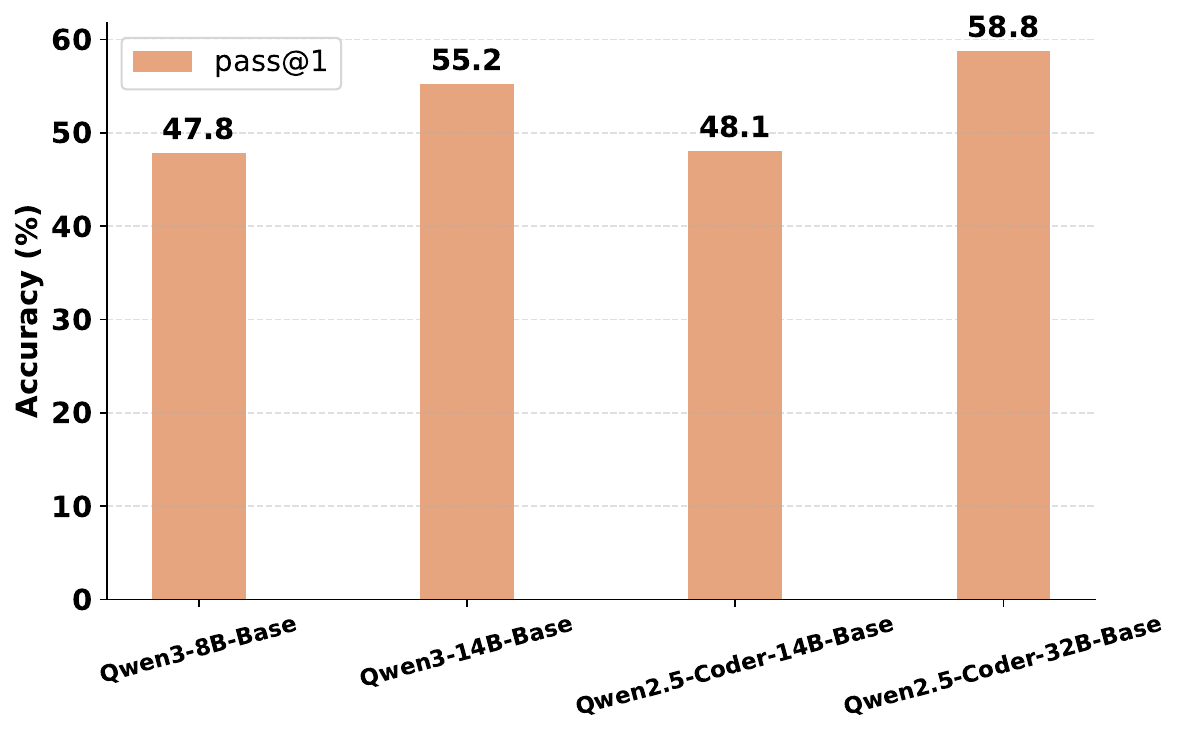}
        \caption{leann}
        \label{fig:leann}
    \end{subfigure}
    \caption{Generality of Different Models. \ref{fig:reaction} shows performance on reaction and \ref{fig:leann} on leann.}
    \vspace{-10pt}
    \label{fig:overview}
\end{figure*}

% \vspace{-10pt}
\subsection{RQ4: Application in Real-World Enterprise Scenarios}
\label{sec:rq4}

\parabf{Design.}
Section~\ref{sec:rq3} demonstrates the potential of \ourapproach for deployment in realistic settings. To further evaluate its effectiveness in real-world enterprise scenarios, where models are typically required to handle multiple codebases across multiple programming languages simultaneously, we conduct an additional experiment in which knowledge from all codebases included in \ourbenchmark is injected into a single model.
We then evaluate this unified model on each individual codebase to assess its ability to retain and apply knowledge across heterogeneous codebases. 
For consistency with previous experiments, we adopt \qwenfourteen as the base model in this setting.

\parabf{Result.}
As shown in Table 3, although the performance of \ourapproach under combined training exhibits a slight degradation compared to separate training, the performance drop remains modest, ranging from only 1.1\% to 4.3\% across different codebases. This behavior is expected and can be attributed to the well-known trade-off and interference effects that commonly arise when training on mixed-domain data. 

Despite this minor degradation, models trained with \ourapproach under the combined setting consistently outperform all baseline methods across every target codebase, achieving an average improvement of 4.7\% to 23.6\% in pass@1. This demonstrates that \ourapproach is robust to cross-codebases and cross-language knowledge integration, and does not rely on isolated, single-codebase adaptation.

\finding{\ourapproach remains effective in real-world industrial scenarios involving multiple programming languages and multiple codebases, achieving a pass@1 of 36.0\% on \ourbenchmark and surparssing all baselines by 4.7\% to 23.6\%}
% \vspace{-10pt}
\subsection{RQ5: Configuration}
\label{sec:rq5}

\parabf{Design.}
We conduct comprehensive experiments to investigate how different configurations under \ourapproach affect model performance. Specifically, we explore:
(1) the impact of varying the number of CPT epochs under a fixed SFT configuration;
(2) the impact of varying the number of SFT epochs under a fixed CPT configuration; and
(3) the effect of different scales of compositional API reasoning data, which contributes most significantly to the model’s performance on unseen codebases.

Due to computational constraints, we select one representative C++ codebase and one Python codebase for these experiments. Since the overall scale of compositional API reasoning data varies across codebases, the staged data sizes used in the experiments are not strictly identical. We adopt \qwenfourteen as the base model for all settings.

\begin{figure*}[b]
    \centering
    \captionsetup[subfigure]{skip=3pt}
    % 统一设置subfigure的底部间距，避免隐式间距差异
    % \setlength{\subfigbottomskip}{0pt}
    % \setlength{\captionsetup}{3pt} % 统一caption与图片的间距
    
    % ---------- 第一类: hexi ----------
    \begin{minipage}[t]{0.32\textwidth}
        \vspace{-5pt} % 关键：重置minipage的顶部对齐基准
        \centering
        \begin{subfigure}[t]{\linewidth} % 显式设置subfigure顶部对齐
            \centering
            \includegraphics[width=\linewidth]{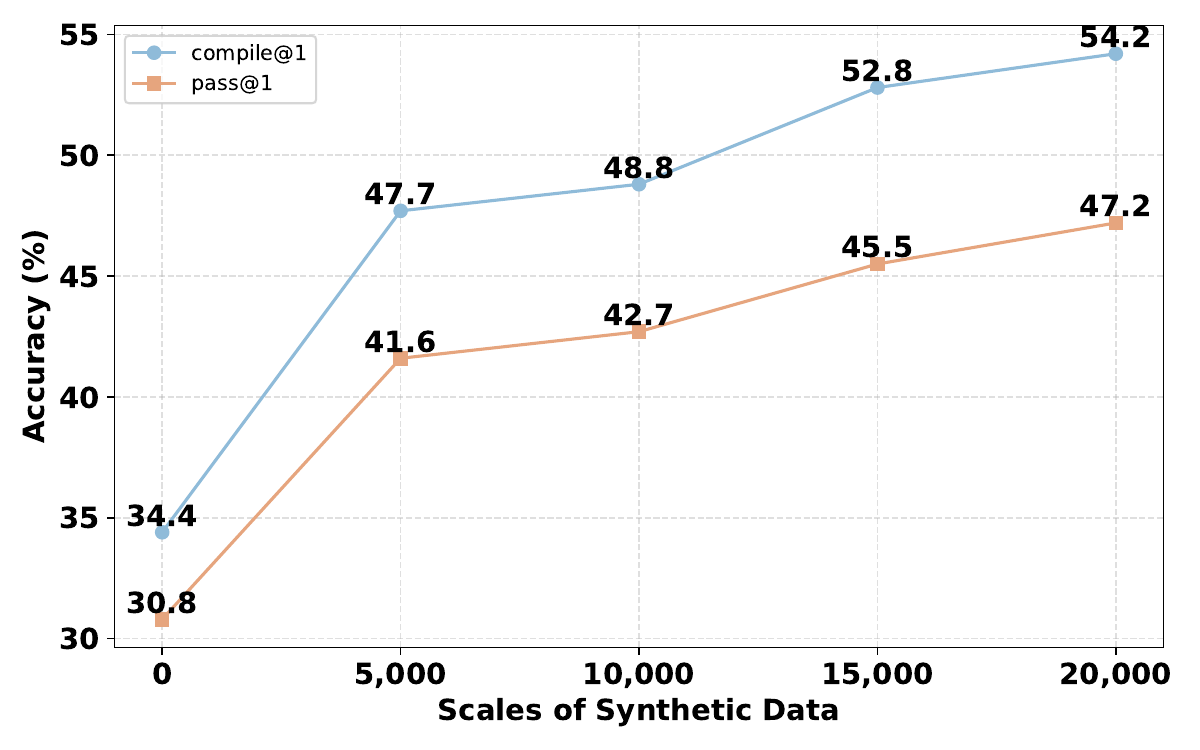}
            \caption{Scale-Hexi}
            \label{fig:hexi_scale}
        \end{subfigure}
        \vspace{0.3cm} % 统一的图片间距
        \begin{subfigure}[t]{\linewidth}
            \centering
            \includegraphics[width=\linewidth]{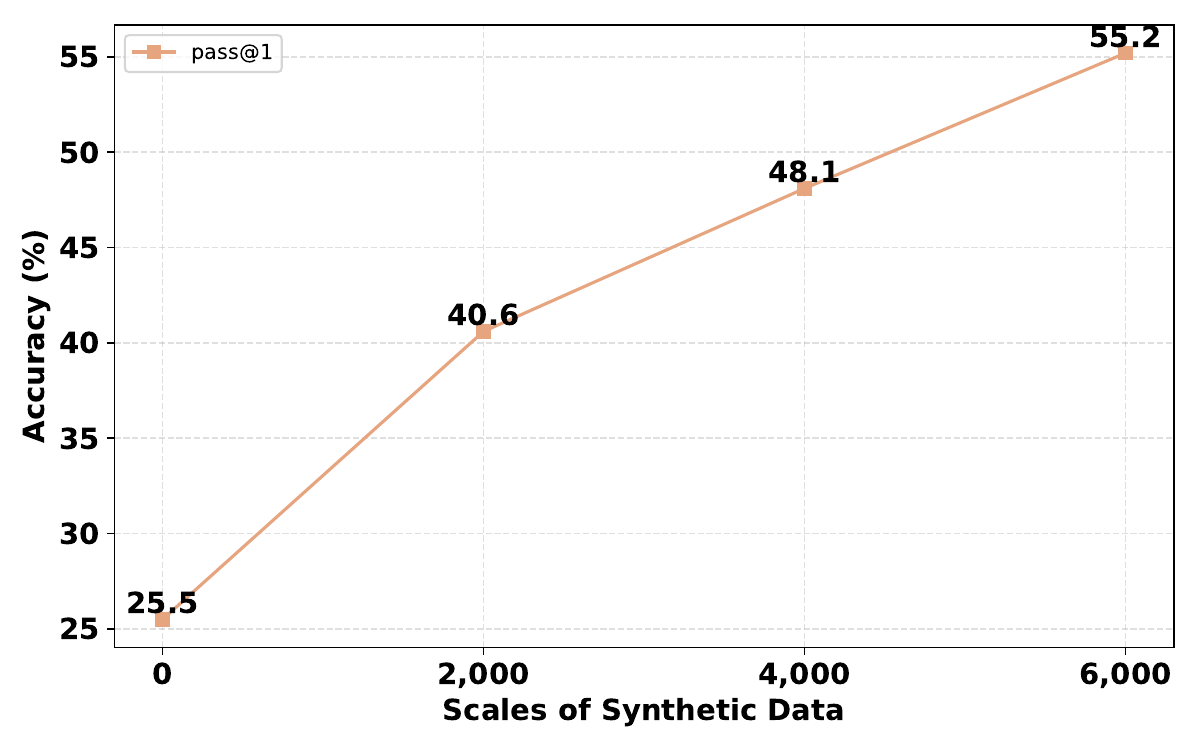}
            \caption{Scale-leann}
            \label{fig:leann_scale}
        \end{subfigure}
        \vspace{-10pt} % 统一caption与最后一张图的间距
        \caption{{Model performance under different synthetic data scales. Base model is \qwenfourteen. \ref{fig:hexi_scale} is model's performance under different scales of synthetic data on reaction, \ref{fig:leann_scale} is model's performance under different scales of synthetic data on leann.}} % \smash消除caption换行的高度影响
        \label{fig:different_scale}
    \end{minipage}
    \hfill
    % ---------- 第二类: leann ----------
    \begin{minipage}[t]{0.32\textwidth}
        \vspace{-5pt} % 关键：和第一列保持一致的顶部基准
        \centering
        \begin{subfigure}[t]{\linewidth}
            \centering
            \includegraphics[width=\linewidth]{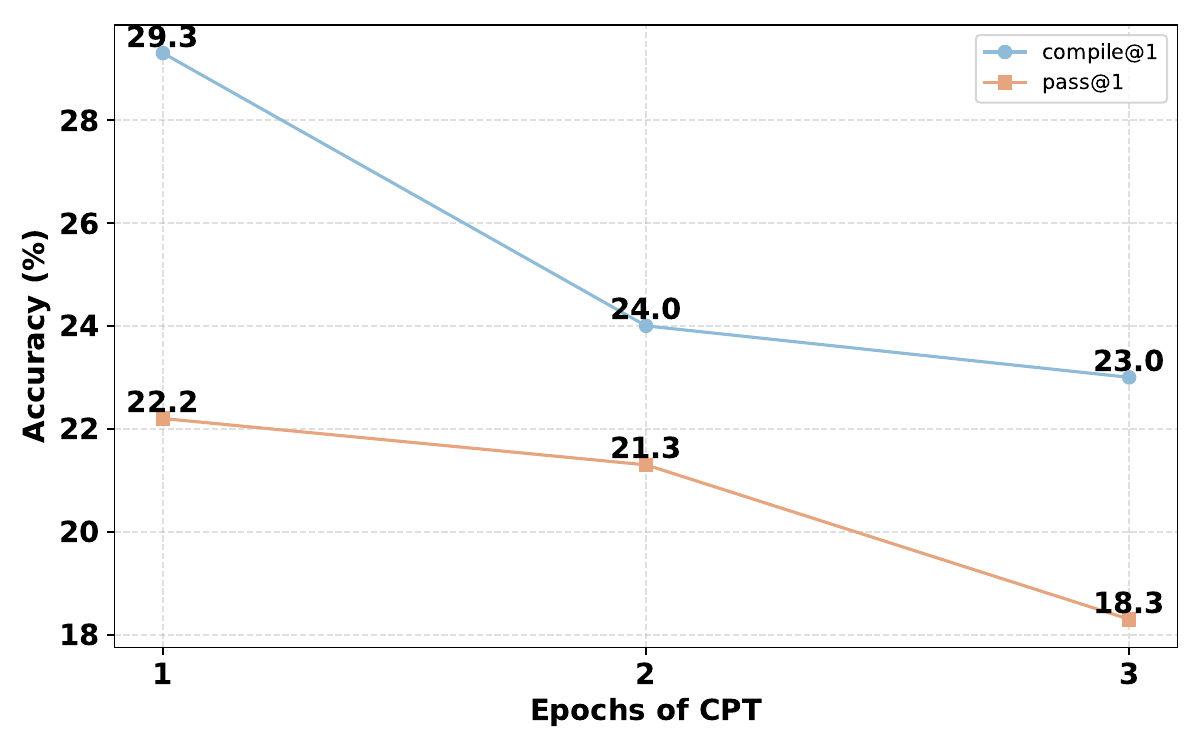}
            \caption{CPT-sqlgen}
            \label{fig:CPT_sqlgen}
        \end{subfigure}
        \vspace{0.3cm} % 统一间距
        \begin{subfigure}[t]{\linewidth}
            \centering
            \includegraphics[width=\linewidth]{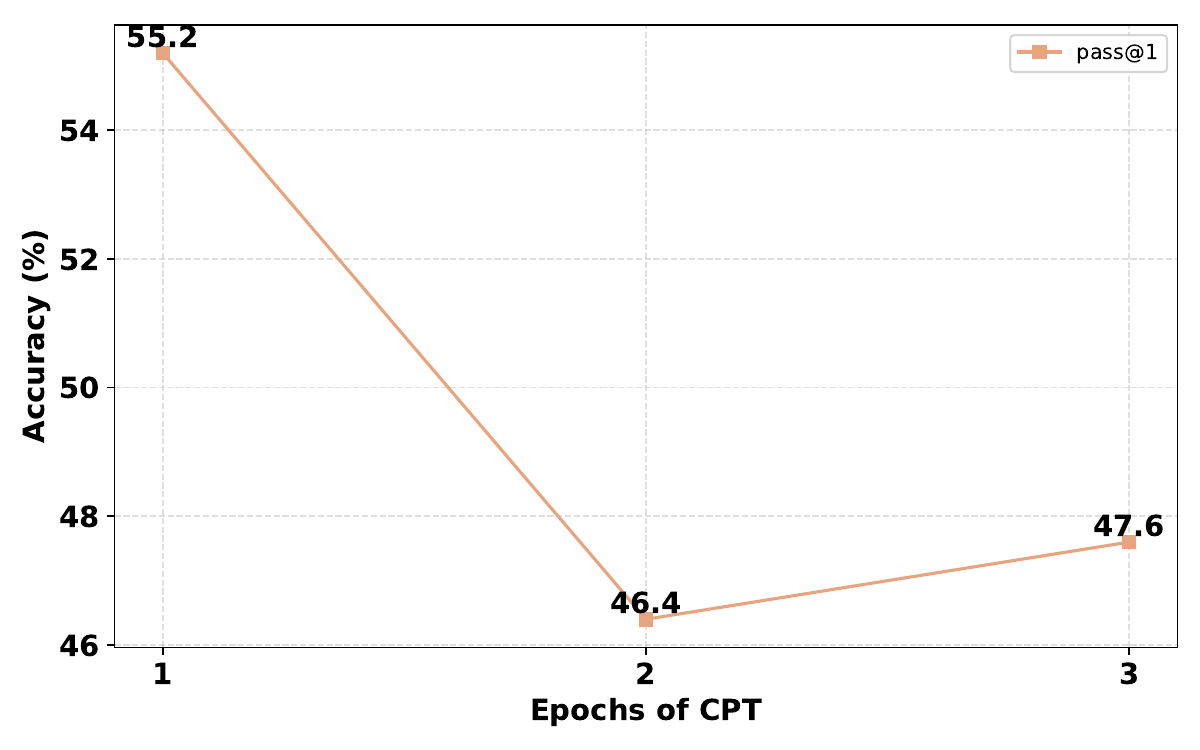}
            \caption{CPT-leann}
            \label{fig:CPT_leann}
        \end{subfigure}
        \vspace{-10pt} % 统一间距
        \caption{{Model performance under different CPT epochs. Base model is \qwenfourteen. \ref{fig:CPT_sqlgen} is Different Model's Performance on sqlgen, \ref{fig:CPT_leann} is Different Model's Performance on leann.}}
        \label{fig:CPT_different_epoch}
    \end{minipage}
    \hfill
    % ---------- 第三类: sqlgen ----------
    \begin{minipage}[t]{0.32\textwidth}
        \vspace{-5pt} % 关键：三列统一顶部基准
        \centering
        \begin{subfigure}[t]{\linewidth}
            \centering
            \includegraphics[width=\linewidth]{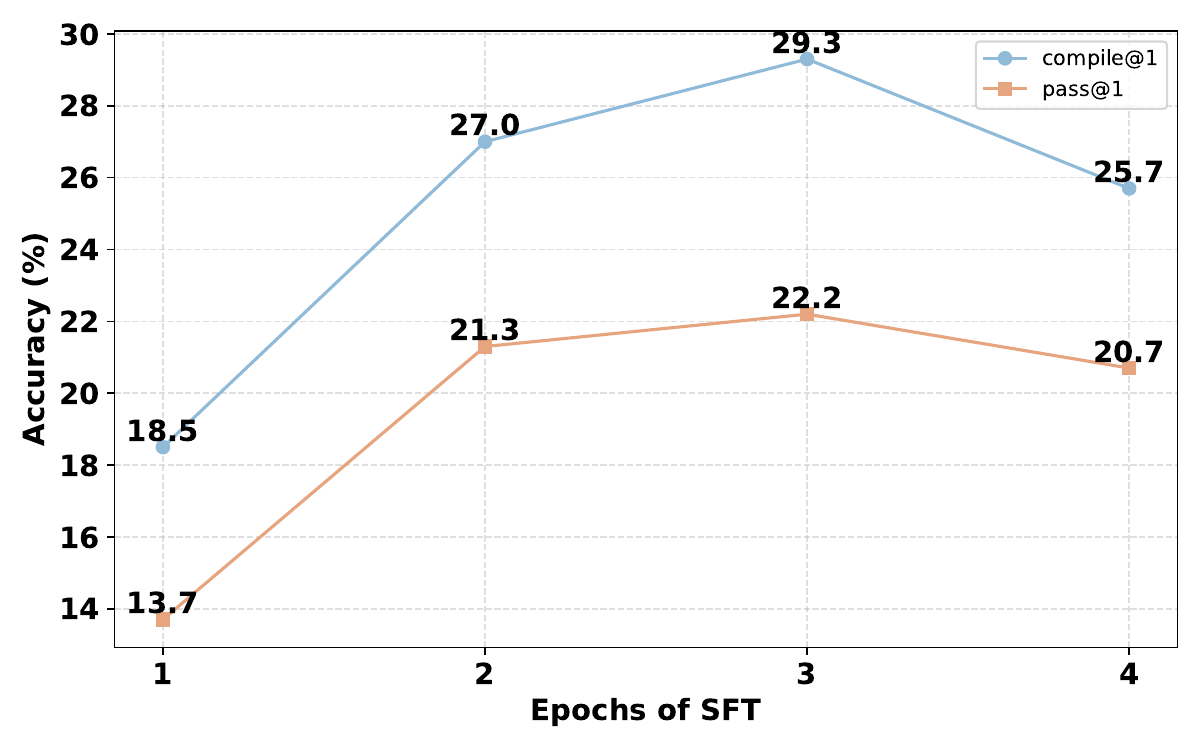}
            \caption{SFT-sqlgen}
            \label{fig:SFT_sqlgen}
        \end{subfigure}
        \vspace{0.3cm} % 统一间距
        \begin{subfigure}[t]{\linewidth}
            \centering
            \includegraphics[width=\linewidth]{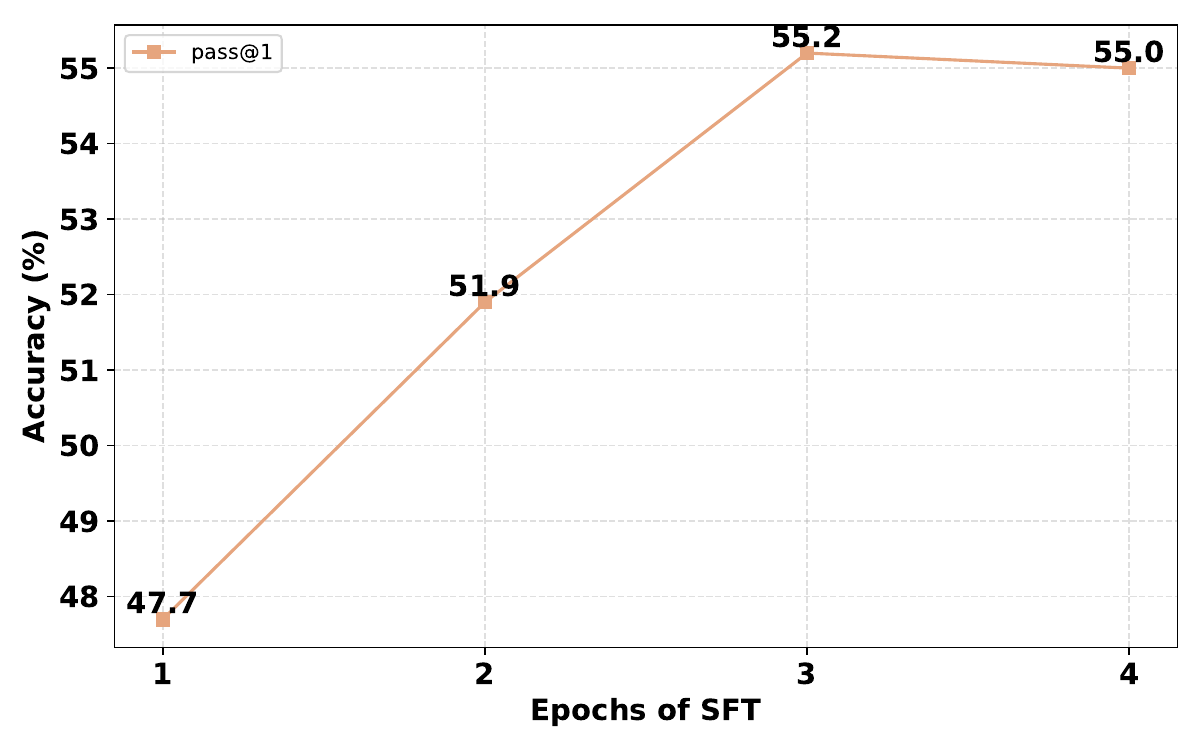}
            \caption{SFT-leann}
            \label{fig:SFT_leann}
        \end{subfigure}
        \vspace{-10pt} % 统一间距
        \caption{{Model performance under different SFT epochs. Base model is \qwenfourteen. \ref{fig:SFT_sqlgen} is Different model's Performance on sqlgen, \ref{fig:SFT_leann} is Different model's Performance on leann}}
        \label{fig:SFT_different_epoch}
    \end{minipage}

\end{figure*}

\parabf{Result.}
As shown in Figures \ref{fig:hexi_scale} and \ref{fig:leann_scale}, the performance of model exhibits a consistent upward trend as the volume of the compositional API reasoning data increases on both codebases, demonstrating the scalability of our proposed data synthesis method. It's important to note that even with the smallest amount of injected knowledge, performance on the Hexi and LEANN codebases improves substantially, from 30.8\% to 41.6\% and from 25.5\% to 40.6\%, respectively. This highlights the high knowledge density of the compositional API reasoning data.

Figures \ref{fig:CPT_sqlgen} and \ref{fig:CPT_leann} further show that the optimal number of epochs for the CPT stage is 1. Increasing the number of CPT epochs beyond this point leads to gradual overfitting. A similar trend is observed in the SFT stage as shown in Figures \ref{fig:SFT_sqlgen} and \ref{fig:SFT_leann}, although the optimal epoch setting differs slightly between CPT and SFT.

The discrepancy between the optimal epoch numbers for CPT and SFT can be attributed to the nature of the training data. While CPT focuses primarily on learning structural and dependency-level information from source code, SFT involves application-oriented data, which requires more training iterations for the model to internalize complex usage patterns and implicit functional relationships.

\finding{Our data synthesis approach demonstrates strong scalability, with model performance improving consistently as the data scale increases. Moreover, compared to source-code-only data, application-level data requires more training epochs to effectively capture its richer relational structure and more complex usage semantics.}

\vspace{-5pt}
\section{Threats To Validity}
\parabf{Threats in Language Coverage.}
Our evaluation focuses on only two programming languages, C++ and Python, and does not include other languages. This limited language coverage may pose a threat to the generalizability of our findings. To further examine the applicability of \ourapproach, we plan to extend our experiments to a broader range of programming languages and development environments in future work, thereby more thoroughly validating the robustness and generalization capability of our method.

\parabf{Threats in Training Configuration.}
Due to computational resource constraints, we were unable to exhaustively explore all possible training configurations during the training process. As a result, the configurations adopted in our experiments may not represent the optimal settings. In future work, we plan to investigate optimal training configurations for unseen codebase domains through both theoretical analysis and more extensive empirical exploration.
\section{Related Work}

\label{sec:related}

\subsection{Data Synthesis for Code LLMs}
Data synthesis techniques aim to automatically construct high-quality instruction–response pairs. Early methods relied on template-based prompting to generate synthetic training data \cite{zhang2026instructionsurvey}, among which Self-Instruct \cite{selfinstruct} is a seminal approach. It iteratively expands an instruction set by bootstrapping from a small number of human-written seeds using an LLM’s own generation capability. This paradigm was later extended to the code domain by Code Alpaca \cite{codealpaca}, which used GPT-3.5 to produce 20K instruction–code pairs. Building on this idea, Evol-Instruct \cite{luo2023wizardcoder} proposed instruction evolution to increase complexity and diversity via in-depth and in-breadth transformations. More recently, OSS-Instruct \cite{wei2023magicoderOSSInstruct} leveraged randomly sampled open-source code snippets as inspiration, prompting LLMs to synthesize diverse programming tasks.

However, most existing data synthesis techniques rely on pre-existing usage-oriented code from the target codebase and are ineffective when only source code is available. To address this limitation, \ourapproach starts directly from source code, performs systematic code analysis to construct a code graph, and then synthesizes the training data base on the code graph.

\subsection{Unseen Domain Code Generation}

Different from generic code generation, unseen-domain code generation requires LLMs to understand domain-specific logic, private APIs, and internal frameworks. Most existing approaches rely on Retrieval-Augmented Generation (RAG), which augments general-purpose code LLMs with externally retrieved knowledge at inference time without updating model parameters \cite{wang2025coderag}. A key challenge lies in effectively leveraging private-library APIs. Prior work~\cite{zan2022language,DomCoder,APICoder,zhang2023repocoder} commonly adopts multi-stage pipelines that retrieve relevant APIs and incorporate them as contextual inputs for code generation.
In practice, domain-specific code generation has been explored across industries. MedCoder~\cite{medcoder} applies LLM-based extraction and retrieval for automated ICD coding; Koziolek et al.~\cite{controlcode} generate IEC 61131-3 ST control code via RAG; AnalogCoder~\cite{analogcoder} focuses on analog circuit design through prompt engineering, while VerilogCoder~\cite{verilogcoder} integrates graph-based planning and AST–waveform tracking for Verilog generation and verification.

However, most existing methods that rely on retrieval-augmented generation are insufficient for enabling models to fully understand the intrinsic relationships among different components of a codebase. Our work synthesizes high-quality architectural and application-level data grounded in the target codebase and leverages post-training to effectively enhance models' understanding and practical usage of the target codebase.

\section{Conclusion}

In this work, we propose \ourapproach, a two-stage training framework that performs reasoning-aware data synthesis based on a code relation graph constructed from unseen codebases. \ourapproach first parses the source code of unseen codebase to build the code graph, then performs continued pretraining (CPT) using file-level dependency data, and finally performs supervised fine-tuning (SFT) with three types of synthesized data augmented with explicit reasoning traces.

We also conduct a new code generation benchmark, \ourbenchmark, targeting unseen codebases and perform comprehensive experiments to demonstrate that \ourapproach consistently outperforms existing baselines across multiple codebases and two programming languages. Furthermore, our method exhibits strong generality: across codebases in different programming languages, models of varying parameter scales, and diverse model architectures, \ourapproach effectively improves models' understanding and application capabilities in unseen codebases domains compared to existing data synthesis baselines, achieving performance gains ranging from 7.2\% to 26.1\% on \ourbenchmark.
We also simulate realistic enterprise scenarios involving multiple languages and multiple target codebase, and show that our approach remains effective in real-world software development settings, achieving a pass@1 of 36.0\% on \ourbenchmark and still surparssing all baselines by 4.7\% to 23.6\%

\section{Data availability}
To facilitate the replication study, we have released our data and benchmark at: ~\url{https://github.com/ooggss/Unseen-Codebases-Domain-Data-Synthesis-and-Training-Based-on-Code-Graphs}.

\bibliographystyle{ACM-Reference-Format}
\bibliography{ref}

\end{document}